\documentclass[pre,
  aps,
  a4paper,
  english,
  showpacs,
  showkeys,
  reprint,
  twocolumn,
  superscriptaddress]{revtex4-1}
\usepackage[T1]{fontenc}
\usepackage[utf8]{inputenc}
\usepackage{amsmath,amsthm,amssymb,graphicx,subfigure,refstyle}
\usepackage{bm}
\usepackage{array,natbib,caption,amsfonts}
\usepackage{mathtools}
 \usepackage{lipsum}
\usepackage{capt-of,verbatim,url}
\usepackage[normalem]{ulem}
\usepackage[
  citecolor=blue,
  colorlinks,
  linkcolor=blue,
  urlcolor=blue,
]{hyperref}
\usepackage[dvipsnames]{xcolor}
\usepackage{hyperref}
\newcommand{\SC}[1]{\textcolor{black}{#1}}

\begin{document}
\newpage
\title{Suppressing birhythmicity by parametrically modulating nonlinearity in limit cycle oscillators}

\author{Sandip Saha}
\email{sandipsaha@bose.res.in} 
\affiliation{S. N. Bose National Centre for Basic Sciences, Block-JD, 
Sector-III, Salt Lake, Kolkata-700106, India}
\author{Sagar Chakraborty}
\affiliation{Department of Physics, Indian Institute of Technology Kanpur, Kanpur, Uttar Pradesh 208016, India}
\email{sagarc@iitk.ac.in} 
\author{Gautam Gangopadhyay}
\email{gautam@bose.res.in} 
\affiliation{S. N. Bose National Centre for Basic Sciences, Block-JD, 
Sector-III, Salt Lake, Kolkata-700106, India}
\begin{abstract}

Multirhythmicity, a form of multistability, in an oscillator is an intriguing phenomenon found across many branches of science. From an application point of view, while the multirhythmicity is sometimes desirable as it presents us with many possible coexisting stable oscillatory states to tap into, it can also be a nuisance because a random perturbation may make the system settle onto an unwanted stable state. Consequently, it is not surprising that there are many natural and artificial mechanisms available that can control the multirhythmicity. \textcolor{black}{We propose in this paper the idea of incorporating parametric (periodic) modulation of the nonlinear damping in the limit cycle oscillators with a view to exciting resonance and antiresonance responses at particular angular driving frequencies, and controlling the resulting birhythmicity by changing the amplitude of the modulation.} To this end, we employ analytical (perturbative) and numerical techniques on the van der Pol oscillator---a paradigmatic limit cycle system---having additional position dependent time delay term and its modified autonomous birhythmic version. We also bring the fact to the fore that introduction of delay---a commonly adopted method of controlling multirhythmicity---in such a system can sometimes bring forth unwanted birhythmicity; and interestingly, our method of controlling birhythmicity through periodic modulation can suppress such a delay induced birhythmic response.
\end{abstract}


\maketitle
\section{Introduction}
\label{penvo}
Since Faraday's observation~\cite{Faraday_1831_paper} of parametric oscillations as surface waves in a wine glass tapped rhythmically, almost two centuries have passed and over the years, it has been realized that the phenomenon of parametric oscillations is literally omnipresent~\cite{Nayfeh_1995_Nonlinear, Marhic_2008_Fiber} in physical, chemical, biological, and engineering systems. Parametric oscillations are essentially effected by periodically varying a parameter of an oscillator which, thus, is aptly called a parametric oscillator. The simplest textbook example with wide range of practical applications is the Mathieu oscillator~\cite{MathieuMmoireSL} where the natural frequency of a simple harmonic oscillator is varied sinusoidally and the interesting phenomenon of parametric resonance~\cite{landau_book} is observed. The effect of additional nonlinearity in the Mathieu oscillator has also been extensively investigated, e.g, in Mathieu--Duffing~\cite{Esmailzadeh1997,PhysRevLett.116.044102}, Mahtieu--van-der-Pol~\cite{Lakrad2005,momeni,Veerman2009}, and Mathieu--van-der-Pol--Duffing~\cite{JiaoruiLi2008,Belhaq2008,Pandey2007,doi:10.1063/1.4938419} oscillators. However, only rather recently, the effect of periodically modulating the nonlinearity in a limit cycle system, viz., van der Pol oscillator has been investigated~\cite{penvo}. The resulting parametric oscillator, termed PENVO (\textbf{p}arametrically \textbf{e}xcited \textbf{n}onlinearity in the \textbf{v}an der Pol \textbf{o}scillator), along with the standard phenomenon of resonance, exhibits the phenomenon of antiresonance that is said to have occurred if there is a decrease in the amplitude of the limit cycle at a certain frequency of the parametrical drive (cf.~\cite{Ewens_book, Saakian_PRE, 2019_sarkar_ray_PRE}).

In the context of the limit cycle oscillations~\cite{Jenkins-2013-PhysicsReports}, one is readily reminded of the limit cycle systems possessing more than one stable limit cycle. A plethora of such multicycle systems are manifested in biochemical processes~\cite{Goldbeter84,gly1,Morita89, Leloup99, Fuente99,Stich2001, Stich2002,gly2}; one of the simplest of them being a multicycle version of the van der Pol oscillator~\cite{kaiser83,kaiser91,k-y2007c} modelling some biochemical enzymatic reactions. This oscillator has two stable limit cycles (and an unstable limit cycle between them in the corresponding two dimensional phase space) owing to the state dependent damping coefficient that has up to sextic order terms. Consequently, it shows birhythmic behaviour wherein depending on the initial conditions, the long term asymptotic solution of the oscillator corresponds to one of the stable limit cycles that have, in general, different frequencies and amplitudes. Needless to say, birhythmicity is a widely found phenomenon across disciplines (\SC{such as biology~\cite{Decroly1982,Goldbeter84,goldbook,goldbeter2002}, physics~\cite{Kwuimy2015}, chemistry~\cite{Alamgir1983}, and ecology and population dynamics~\cite{Arumugam2017rhythmogenesis}}) because so are the ubiquitous limit cycle oscillations.

Since different initial conditions lead to different solutions for a birhythmic oscillator, the inherent uncertainty in the amplitude and the frequency in the eventually realized stable oscillations can be gotten rid of if the oscillator is somehow made monorhythmic. It is known~\cite{Pisarchik2000,Pisarchik2003,Goswami2008} from the studies on the H\'enon map and rate equations of laser that while a small change of one of the system parameters of a birhythmic oscillator may not in general convert it to a monorhythmic system, an external control in the form of a slow periodic parameter modulation can annihilate one of the coexisting attractors resulting in a monostable oscillatory system. Technically speaking, birhythmicity is a simple type of multistability which, in other words, mean coexistence of different attractors at fixed parameter values in the system. The existence of multistability in diverse systems and the need to control it are elaborately discussed in a review article~\cite{Pisarchik2014} which also reviews various control strategies including their experimental realizations. 

Interestingly, time delay is known to have significant effect on the attractors of a nonlinear system and can also brings forth new ones\SC{~\cite{Cooke_Grossman_82,Banerjee2012design,Banerjee2013FirstOrderChaotic,Biswas2016simple,Park2019}}. For example, even in a relatively simple system like the R\"ossler oscillator, time delayed feedback control~\cite{Balanov2005} induces a large variety of regimes, like tori and new chaotic attractors, nonexistent in the original system; furthermore, the delay modifies the periods and the stabilities of the limit cycles in the system depending on the strength of the feedback and the magnitude of the delay. As another example, we may point out that the direct delayed optoelectronic feedback can suppress hysteresis and bistability in a directly modulated semiconductor laser~\cite{Rajesh2006}. The coexistence of two stable limit cycles with different frequencies in the presence of delayed feedback has been discussed in detail~\cite{Erneux2008} for the van der Pol oscillator and its variants.  Mutlicycle van der Pol oscillator has also been investigated from the point of view of control of birhythmicity using some different forms of time delay~\cite{k-y2007a,k-y2007b,k-y2007c,k-dsr}. 

\textcolor{black}{However, to the best of our knowledge, there has been no investigation into the control of multistability in a parametric oscillator whose parameter, determining the strength of the nonlinear term, is varied. While our study has been somewhat driven by the lack of any such prior investigation and mathematical curiosity, it should be noted that periodic variation of such a parameter is not inconceivable~\cite{2016-ghosh-dsr-PRE}; in fact, it can result in parametric spatiotemporal instability leading to interesting time-periodic stationary patterns in reaction-diffusion systems. Furthermore, it may be worth pointing out that the van der Pol oscillator forms a crucial ingredient in modelling the mechanical resonators based on carbon nanotubes and graphene sheets, where it is known that damping depends on amplitude of the oscillations~\cite{Eichler2011, Singh2020giant}; how the damping coefficients are best modelled is not a completely answered question. Moreover, in principle, the experiments concerned with the graphene-resonators can design time dependent nonlinear damping.} 

In view of the above, it is imperative that an investigation of the PENVO and its relevant extension be carried out and the interplay, if any, between the time-delayed feedback and the parametric forcing be revealed. To this end, in this paper, we first discuss in Sec.~\ref{sec2} how presence of time delayed feedback affects the resonance and the antiresonance in the PENVO. Furthermore, we discuss how the resulting birhythmicity therein is suppressed by tuning the strength of the period modulation. Subsequently, in Sec.~\ref{KBM}, we consider multicycle PENVO---multicycle van der Pol oscillator whose nonlinearity is sinusoidally varying---and argue in detail that it is possible to control birhythmicity in this system as well. Finally, we reiterate the main results of this paper in Sec.~\ref{C}.

\section{PENVO with Delay}
\label{sec2}
\begin{figure}
\includegraphics[width=4cm, height=4cm]{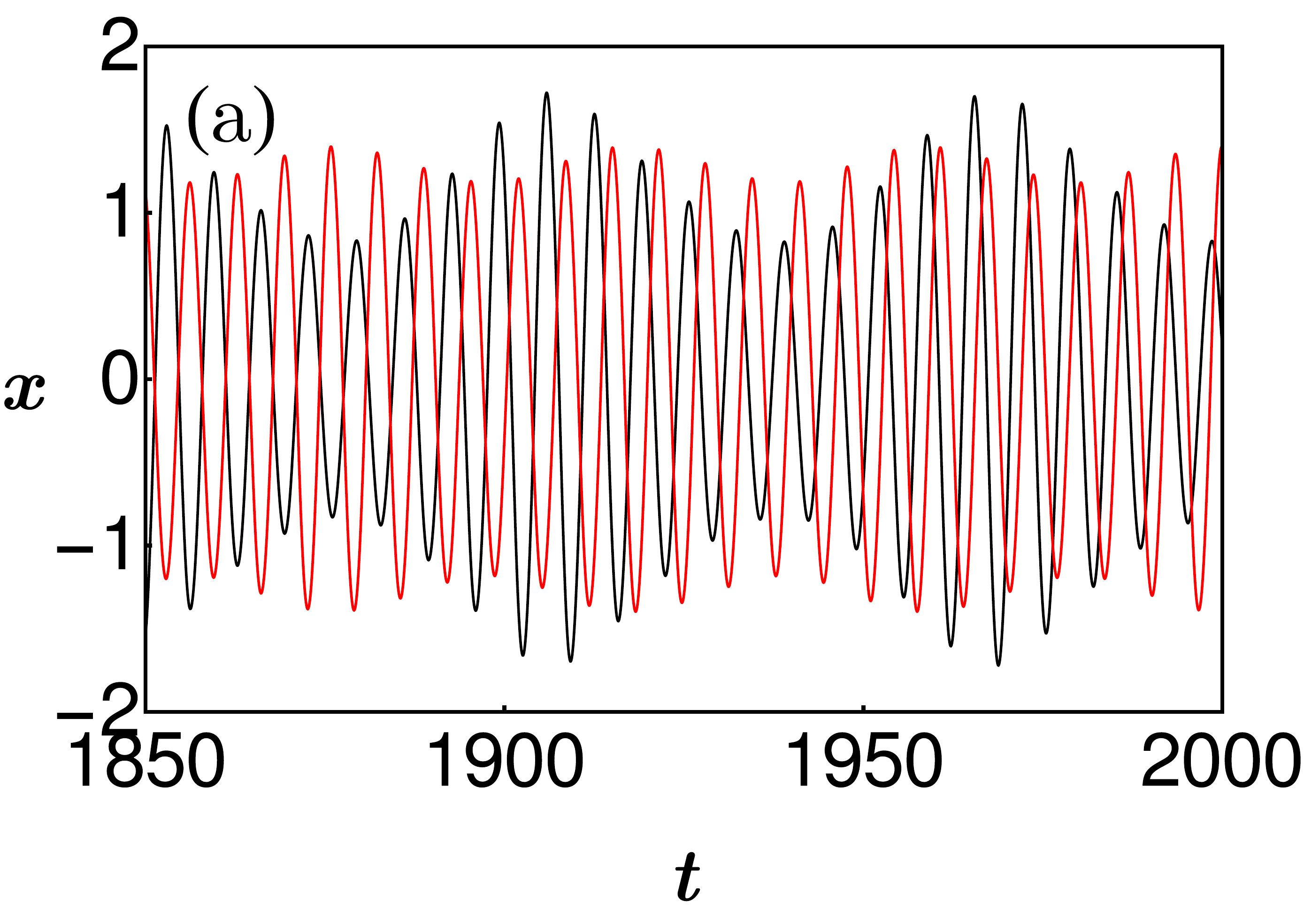}
\includegraphics[width=4cm, height=4cm]{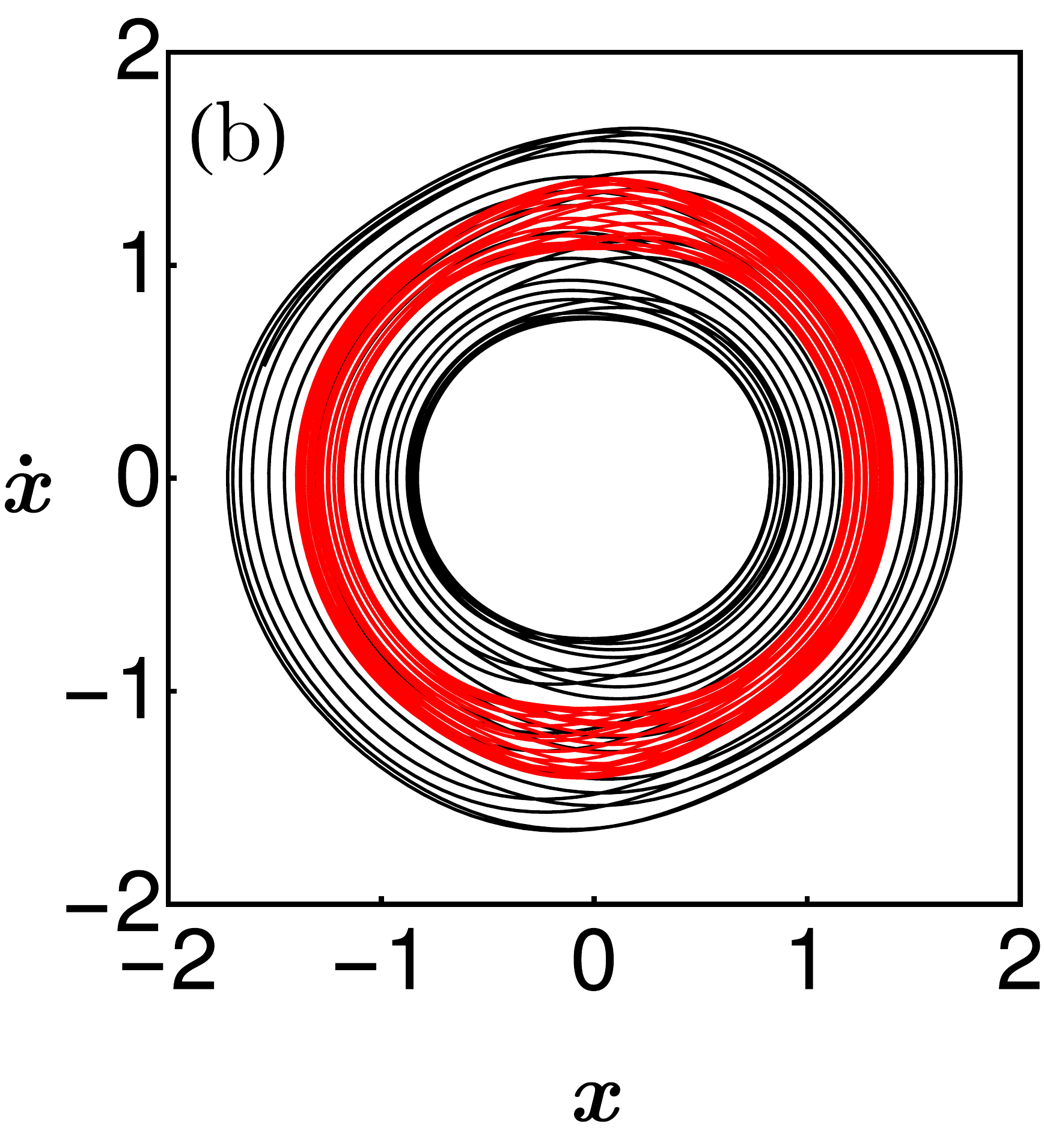}
\caption{\emph{Limit cycles in PENVO with delay have oscillating amplitudes.} We time-evolve Eq.~(\ref{eq:pentdvo}) with $\gamma=1.5,\,K=\mu=0.1,\,\tau=0.623$ for $\Omega=2~{\rm (black)~and}~4~{\rm(red)}$ to arrive at the corresponding time-series plots (subplot a), $x$~vs.~$t$, and phase space plots (subplot b), $\dot{x}$~vs.~$x$.}
\label{fig:delayed vdp} 
\end{figure}
Even a simple harmonic oscillator with its quadratic potential modified so as to have a term that is time delayed, exhibits  nontrivial dynamics. The resulting solutions, including the oscillatory ones, in the weak nonlinear limit can be iteratively extracted using perturbative methods based on the concept of renormalization group~\cite{goto2007renormalization, len3.5}. An extended version of the delayed simple harmonic oscillator, that possesses limit cycle, has also been analyzed~\cite{powerlaw} using the Krylov--Bogoliubov method~\cite{kbbook, jkb2007}. Motivated by these results, we now consider the PENVO with a time delay term as follows:
\begin{eqnarray}
 \ddot{x} + \mu [1+\gamma \cos(\Omega t)] (x^2 -1) \dot{x}  + x - K x(t-\tau)=0,\qquad
\label{eq:pentdvo}
\end{eqnarray}
where  $0<K, \mu \ll 1$; \SC{$\tau<1$}; $\gamma\in\mathbb{R}$; and $\Omega \in \mathbb{R}^+$.

Note that for $K=\gamma=0$, we get back the van der Pol oscillator that in weak nonlinear limit shows stable limit cycle oscillations with amplitude 2. For appropriate non-zero values of $\gamma$ ($K$ still zero), we arrive at the equation for the PENVO~\cite{penvo} that is known to show antiresonance (oscillations with amplitude smaller than 2) and resonance (oscillations with amplitude greater than 2) at $\Omega=2$ and $\Omega=4$ respectively. \emph{Our specific goal in this section is to find out what happens to the resonance and the antiresonance states once the time delay is introduced (i.e., when $K,\gamma\ne0$ and $\Omega=2,4$), and to explore the possible existence of birhythmicity and its control in the system.}
\begin{figure*}
\includegraphics[width=5.6cm, height=5cm]{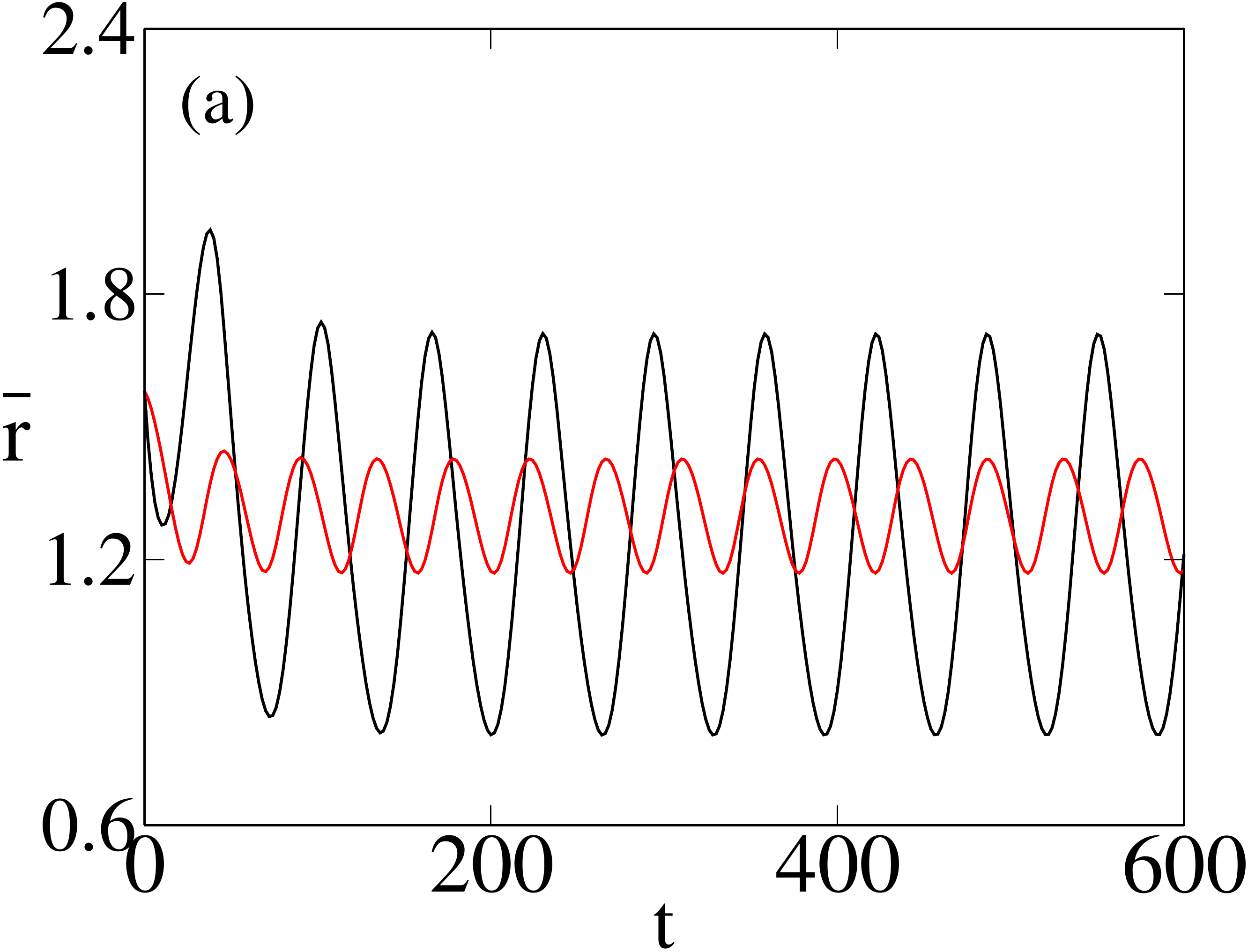}
\includegraphics[width=5.6cm, height=5cm]{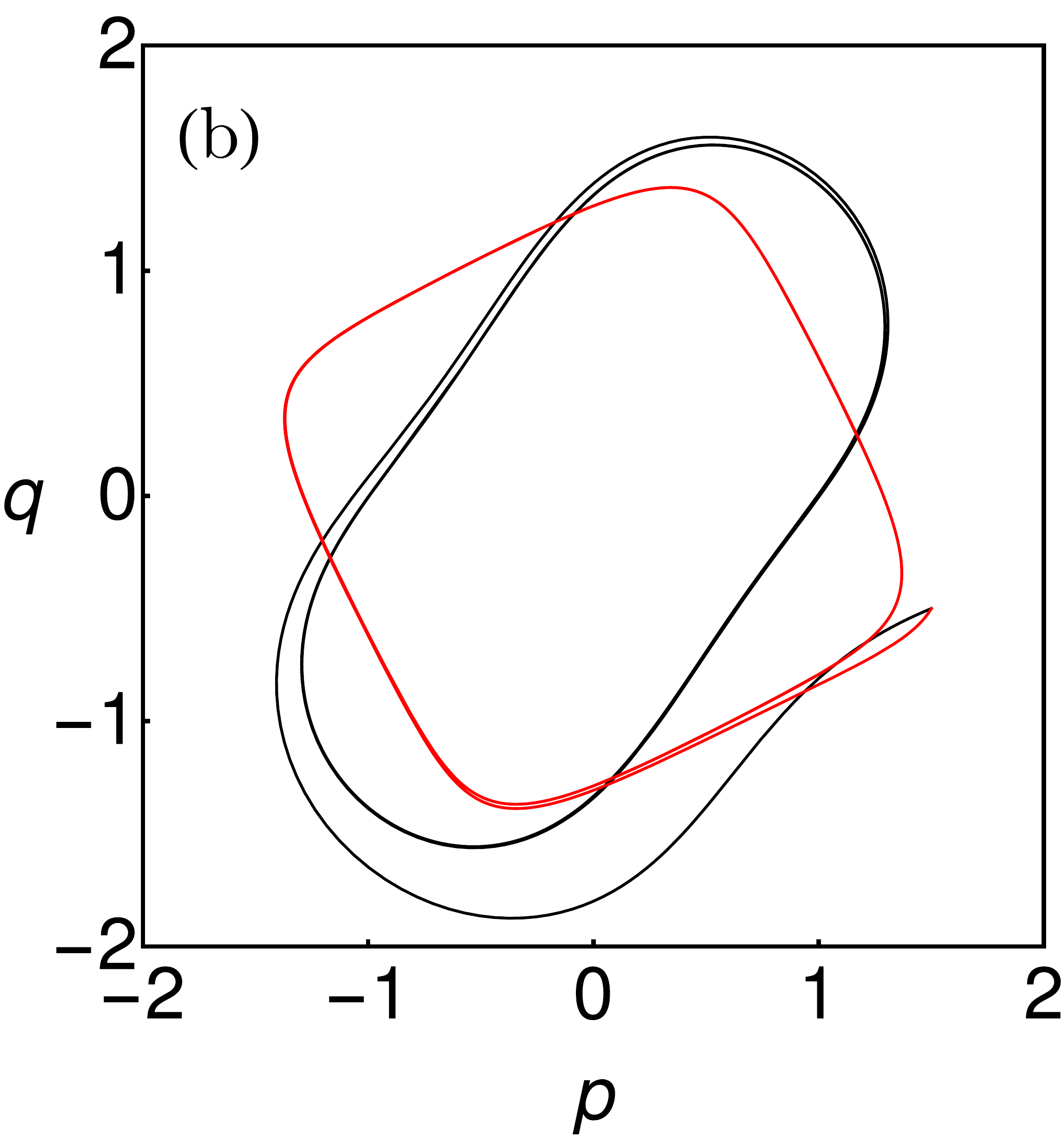}
\includegraphics[width=5.6cm, height=5cm]{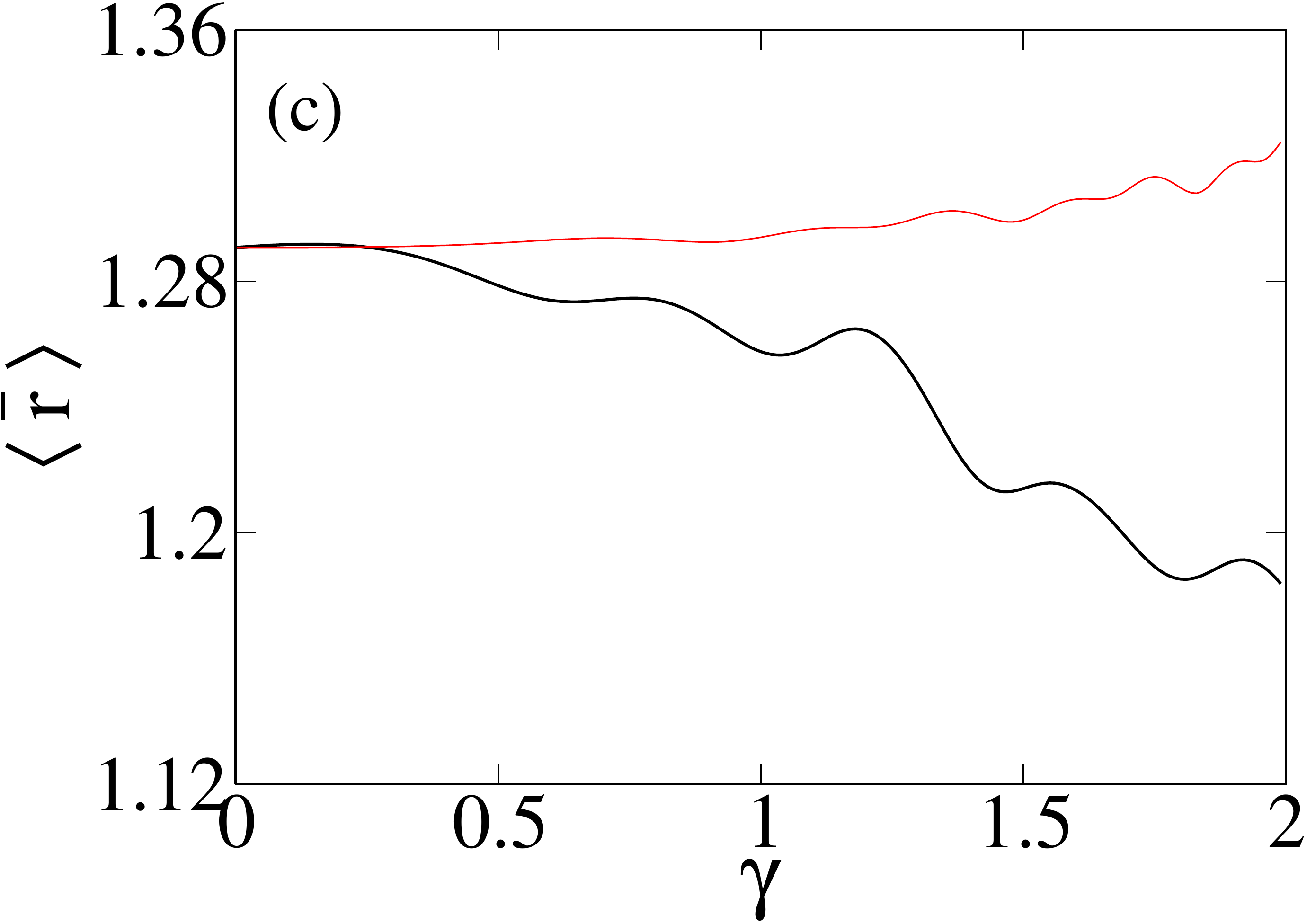}
\caption{\emph{Anti-resonant responses with oscillating amplitudes in PENVO with delay.} This figure panel has been generated by time-evolving Eq.~(\ref{eq:pentdvo}) with $\gamma\in[0,2]$, $K=\mu=0.1,\,\tau=0.623$; and $\Omega=2~{\rm (black)~and}~4~{\rm(red)}$. The time-series, $\overline{r}$~vs.~$t$, (subplot a) depicts oscillating limit cycles in the PENVO with delay and the reason behind the oscillations is best understood as the corresponding non-circular limit cycle attractors in the $p$-$q$ plane (subplot b). While for subplots (a) and (b), $\gamma=1.5$, subplot (c) showcases the variation of the averaged amplitudes with $\gamma$, thus, highlighting the presence of antiresonances $\forall\gamma\in[0,2]$.}
\label{fig:pentdvo_figosr} 
\end{figure*}
\begin{figure*}
\includegraphics[width=4.3cm, height=5cm]{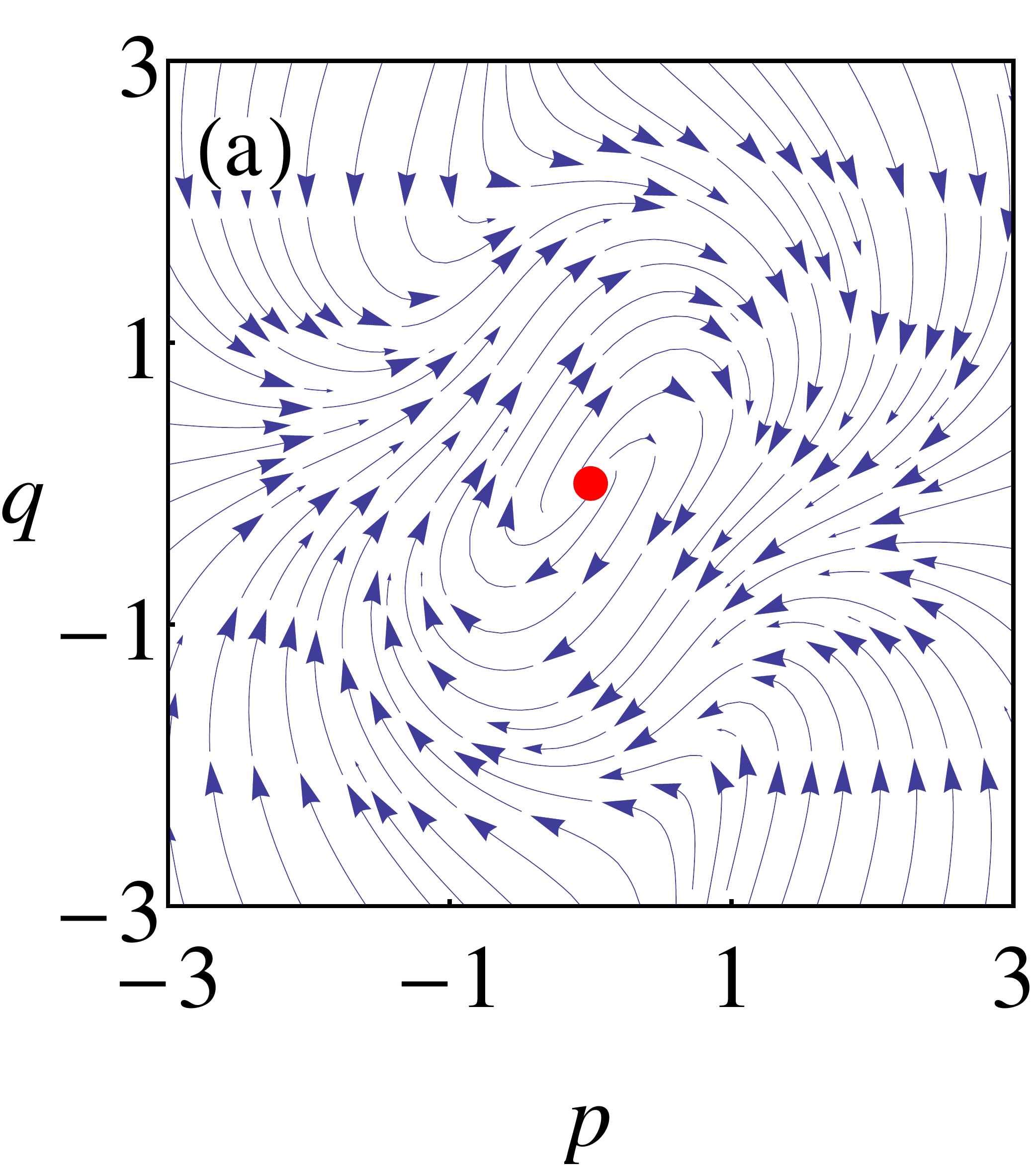}
\includegraphics[width=4.3cm, height=5cm]{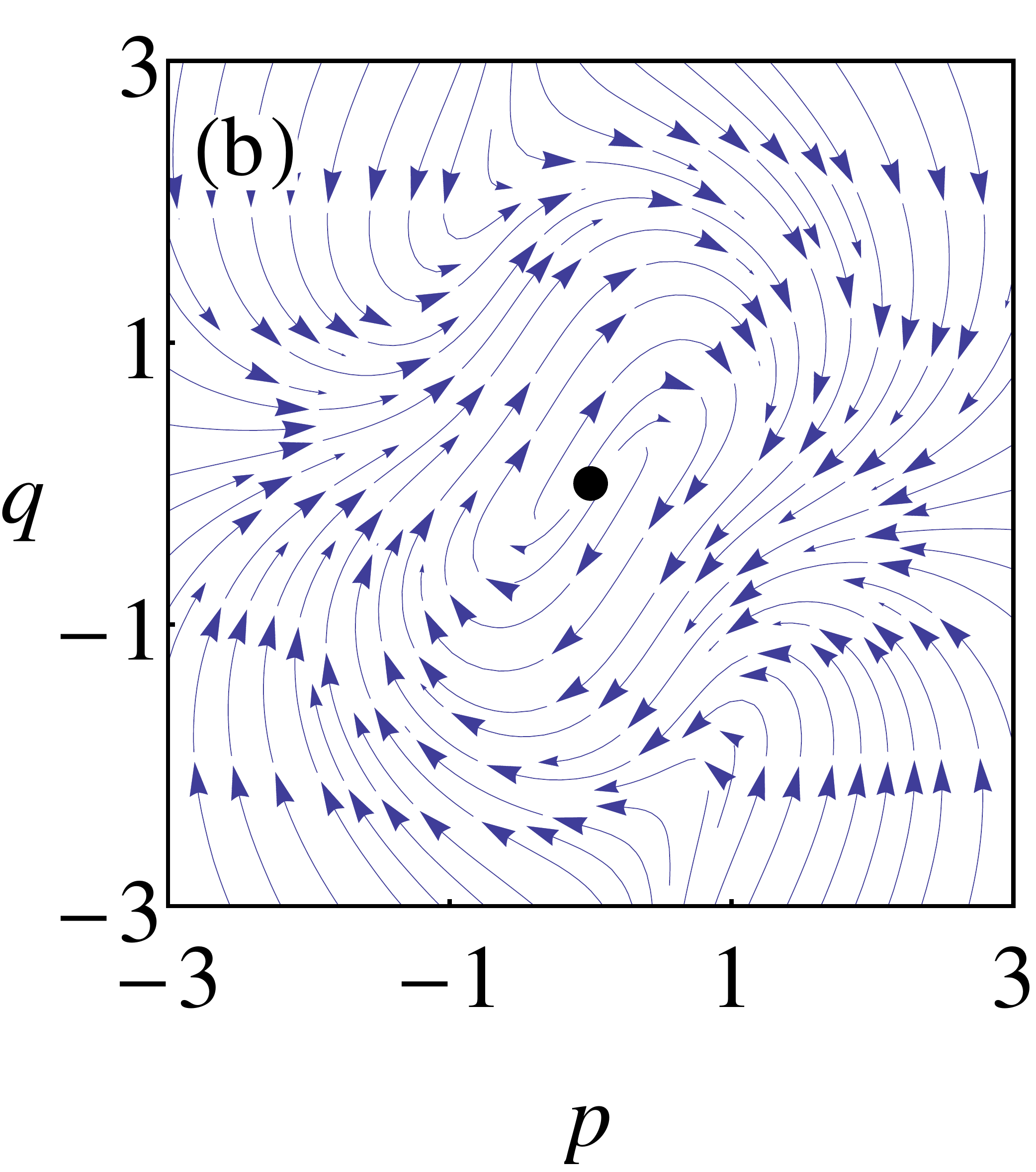}
\includegraphics[width=4.3cm, height=5cm]{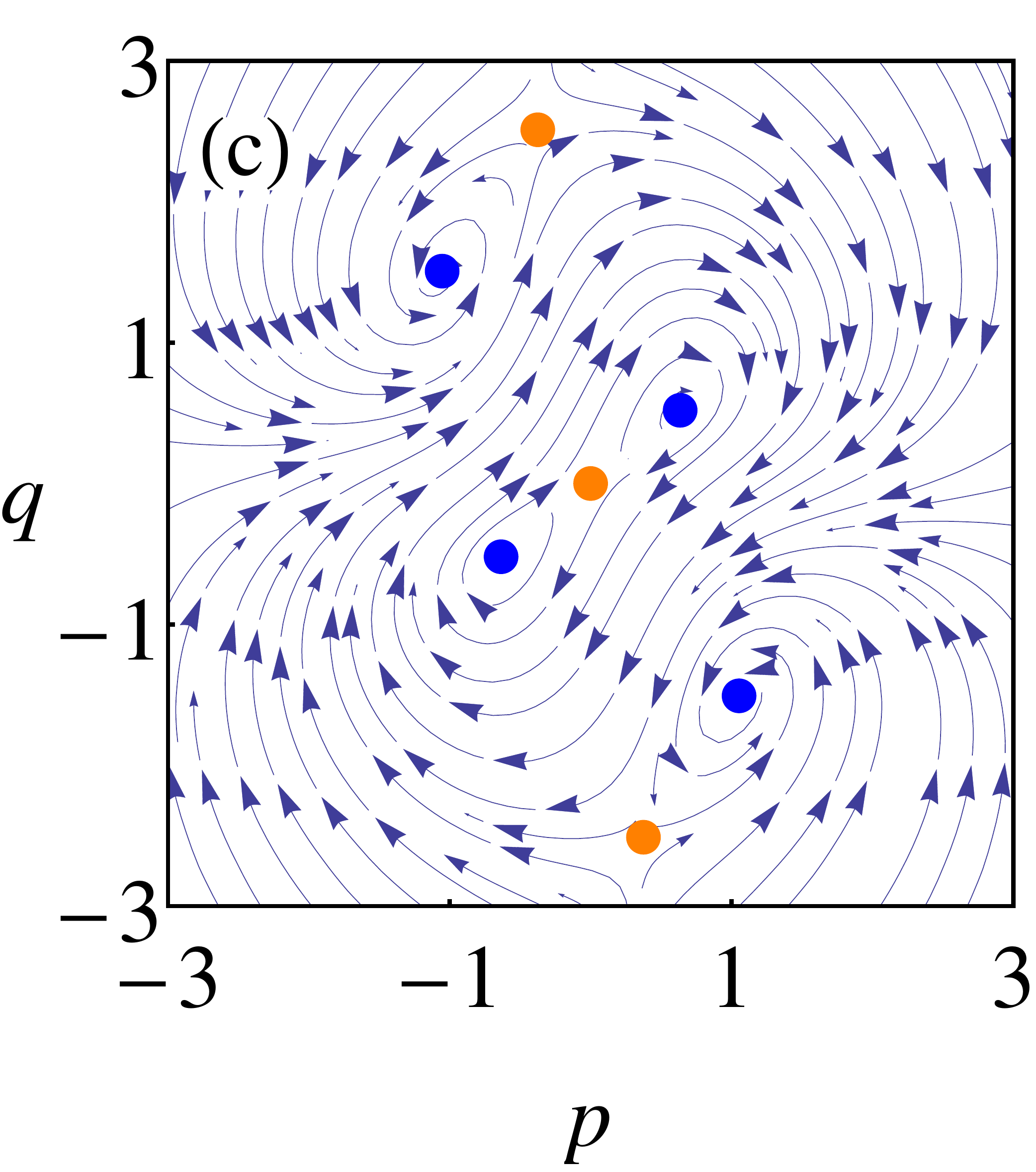}
\includegraphics[width=4.3cm, height=5cm]{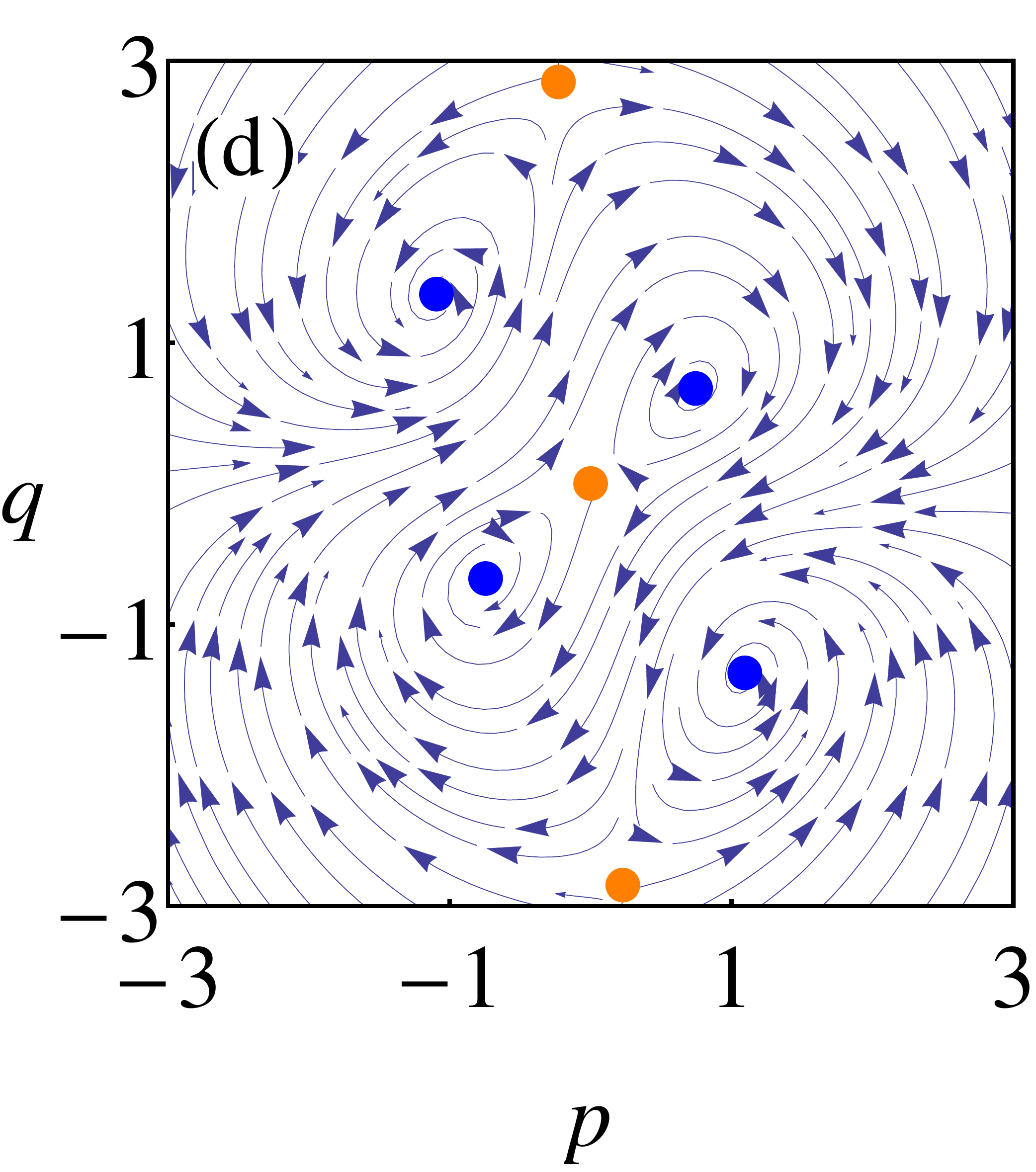}
\caption{\emph{Strength of periodic modulation of nonlinear damping controls delay-induced birhythmicity.} This figure panel of streamline plots depicts repellers [unstable focus (red dot), \SC{unstable node (black dot)} and saddle (orange dot)] and attractors [stable focus (blue dot) and stable limit cycle (around the unstable focus and the unstable  node; not explicitly shown)] in $p$-$q$ space of the PENVO with delay at $\gamma=1.5~({\rm subplot~a}),~\SC{1.7~({\rm subplot~b})},~2.5~({\rm subplot~c}),~\textrm{and}~3.3~({\rm subplot~d}); \,K=\mu=0.1;\,\tau=0.623$; and $\Omega=2$. The stable foci on (approximately) principle diagonal of the figures have same $\sqrt{p^2+q^2}$-value, and so is the case with the stable foci on (approximately) anti-diagonal of the figures. Note how with change in $\gamma$-value, the number of attractors changes from one (limit cycle) to four (foci that have only two distinct $\sqrt{p^2+q^2}$-value).}
\label{fig:stabilityswitching}
\end{figure*}

To begin with we have extensively searched for numerical solutions of Eq.~(\ref{eq:pentdvo}) at different parameter values. In Fig.~\ref{fig:delayed vdp}, we present two particular oscillatory solutions for the cases $\Omega=2$ and $\Omega=4$. We note that the limit cycles have oscillating amplitudes. In order to understand the origin of oscillating amplitude and to discover birhythimicity in the course of our investigation, we employ the Krylov--Bogoliubov method on Eq.~(\ref{eq:pentdvo}). We, thus, make an ansatz: $x(t)=r(t)  \cos ( t+\phi(t))$ where we have adopted polar coordinate, $\left(r, \phi \right) = (\sqrt{x^2 +{\dot{x}^2}}, - t+ \tan^{-1} (- {\dot{x}}/{ x}))$. $r$ and $\phi$ are very slowly varying function of time since we are working under the assumption that $0<\mu\ll1$; we set $r(t) =\overline{r}+O(\mu)$ and
$\phi(t) =\overline{\phi}+O(\mu)$. Here, we have used the definition that average of a function, $f(x,\dot{x})$ (say), over a period $2\pi$ is conveniently denoted as $\overline{f}(t) = ({1}/{2 \pi}) \int_{0}^{{2 \pi}} f(s) ds$. Furthermore, Taylor-expanding $r(t-\tau)$ as $r(t-\tau)=r(t)-\tau \dot{r}(t)=r(t)+O(\mu)$  (since $\dot{r}(t)\sim O(\mu)$), one finally obtains
\begin{subequations}
\begin{eqnarray}
\dot{\overline{r}} &= &-\frac{\overline{r} \left(4 K \sin \tau +\mu  \left(\overline{r}^2-4\right)  \right)}{8  }+A_\Omega (\overline{r},\overline{\phi};\gamma)+O(\mu^2),\nonumber\\ \\
\dot{\overline{\phi}} &= &-\frac{K \cos \tau }{2 }+B_\Omega(\overline{r},\overline{\phi};\gamma)+O(\mu^2),
\end{eqnarray}
\label{eq:pentdvo_amp_ph}
\end{subequations}
where, $O(\mu^2)$ terms can be neglected and $A_\Omega$ and $B_\Omega$ denote the $\gamma$ dependent parts. It is interesting that these two functions' denominators blow up at $\Omega$ equal to $2$ and $4$. We, thus, resort to the L'H\^ospitals' rule to find the functions at $\Omega = 2, 4$:
\begin{subequations}
\begin{eqnarray}
A_{2}(\overline{r},\overline{\phi};\gamma)&=&-\frac{1}{4} \gamma \mu \overline{r}  \cos (2 \overline{\phi} ),\quad\\
B_{2}(\overline{r},\overline{\phi};\gamma)&=&-\frac{1}{8} \gamma  \mu  \sin (2 \overline{\phi} ) \left(\overline{r}^2-2\right);\\
A_{4}(\overline{r},\overline{\phi};\gamma)&=&\frac{1}{16} \gamma  \overline{r}^3 \mu  \cos (4 \overline{\phi} ),\quad\\
B_{4}(\overline{r},\overline{\phi};\gamma)&= &-\frac{1}{16} \gamma  \overline{r}^2 \mu  \sin (4 \overline{\phi} ).
\end{eqnarray}
\end{subequations}
Here the subscripts specify the value of $\Omega$ at which $A_\Omega$ and $B_\Omega$ have been determined.

As an illustration, in Fig.~\ref{fig:pentdvo_figosr}(a), we present $\overline{r}$ as a function of $t$ for both 
$\Omega=2$ and $\Omega=4$ after fixing $\gamma=1.5$, $\tau=0.623$, and $K=\mu=0.1$. The \emph{solutions are oscillatory in sharp contrast to the case of the weakly nonlinear van der Pol oscillator} for which the plot of $\overline{r}$ vs. $t$ would be a horizontal straight line passing through $\overline{r}=2$ at large times. Obviously, it is a little ambiguous to define the resonance and the antiresonance states in terms of the magnitude of the oscillations' amplitude because the amplitude itself is oscillating. Hence for the sake of consistency, to define the resonance and the antiresonance states, we henceforth use the average of the oscillating amplitude. Consequently, in Fig.~\ref{fig:pentdvo_figosr}(c), we plot average of $\overline{r}$ i.e. $\langle \overline{r} \rangle_t$  (after removing enough transients) with $\gamma$ to note that at both $\Omega=2$ and $\Omega=4$ the system shows antiresonance. \SC{Note that \emph{one of the interesting effects of the delay is to suppress the uncontrolled growth of oscillations} (at $\Omega=4$ and as $\gamma\rightarrow2$) {{present in the absence of delay}}. In the PENVO model ({\it i.e.,} $K=0$ in Eq.~\ref{eq:pentdvo})~\cite{penvo}, it has been observed that there exists an antiresonance state at $\Omega=2$ and a resonance (in the form of uncontrolled growth of oscillations) at $\Omega=4$. However, on introducing the position dependent delay, the extended model (Eq.~\ref{eq:pentdvo}) results in suppressing the resonance phenomena to yield an antiresonance state at $\Omega=4$.} 

The oscillations in the amplitudes of the limit cycles is best explained by recasting the equations for $\overline{r}$ and $\overline{\phi}$ in $(p,q)$-plane where $(p,q)=\left( \overline{r} \cos\overline{\phi},\overline{r} \sin\overline{\phi}\right)$ or consequently, $(\overline{r}, \overline{\phi})=( \sqrt{p^2+q^2},\tan ^{-1}({q}/{p}) )$. Substituting these relations in equations~(\ref{eq:pentdvo_amp_ph}), one arrive at the following dynamical flow equations: 
\begin{widetext}
\begin{subequations}
\begin{eqnarray}
 \dot{p}|_{2 }&=&-\frac{K p \sin \tau}{2  }+\frac{K q \cos \tau}{2  }-\frac{\mu  p^3}{8}-\frac{\gamma  \mu  p}{4}+\frac{\mu  p}{2}+\frac{1}{4} \gamma  \mu  p q^2-\frac{1}{8} \mu  p q^2, \\
\dot{q}|_{2 }&=&-\frac{K p \cos \tau }{2  }-\frac{K q \sin \tau  }{2  }-\frac{1}{4} \gamma  \mu  p^2 q-\frac{1}{8} \mu  p^2 q-\frac{\mu  q^3}{8}+\frac{\gamma  \mu  q}{4}+\frac{\mu  q}{2};
\end{eqnarray}
\label{eq:pentdvo_vdp_plane_2}
\end{subequations}
\begin{subequations}
\begin{eqnarray}
\dot{p}|_{4 }&=&-\frac{K p \sin \tau}{2  }+\frac{K q \cos \tau }{2  }+\frac{1}{16} \gamma  \mu  p^3-\frac{\mu  p^3}{8}+\frac{\mu  p}{2}-\frac{3}{16} \gamma  \mu  p q^2-\frac{1}{8} \mu  p q^2,\\
\dot{q}|_{4 }&=&-\frac{K p \cos \tau }{2  }-\frac{K q \sin \tau}{2  }-\frac{3}{16} \gamma  \mu  p^2 q-\frac{1}{8} \mu  p^2 q+\frac{1}{16} \gamma  \mu  q^3-\frac{\mu  q^3}{8}+\frac{\mu  q}{2}.
\end{eqnarray}
\label{eq:pentdvo_vdp_plane_4}
\end{subequations}
\end{widetext}
Here again subscripts $2$ and $4$ refer respectively to the cases corresponding to $\Omega=2$ and $\Omega=4$. Fig.~\ref{fig:pentdvo_figosr}(b) exhibits the limit cycles that are not perfect circles about the origin in $p$-$q$ plane. Thus, it is clear that for either of the cases, the slow variation of the limit cycle amplitude is manifested through the slow variation of the distance of the phase point on the closed trajectory from the origin in $p$-$q$ plane.

\begin{figure}
\includegraphics[width=8cm, height=5.5cm]{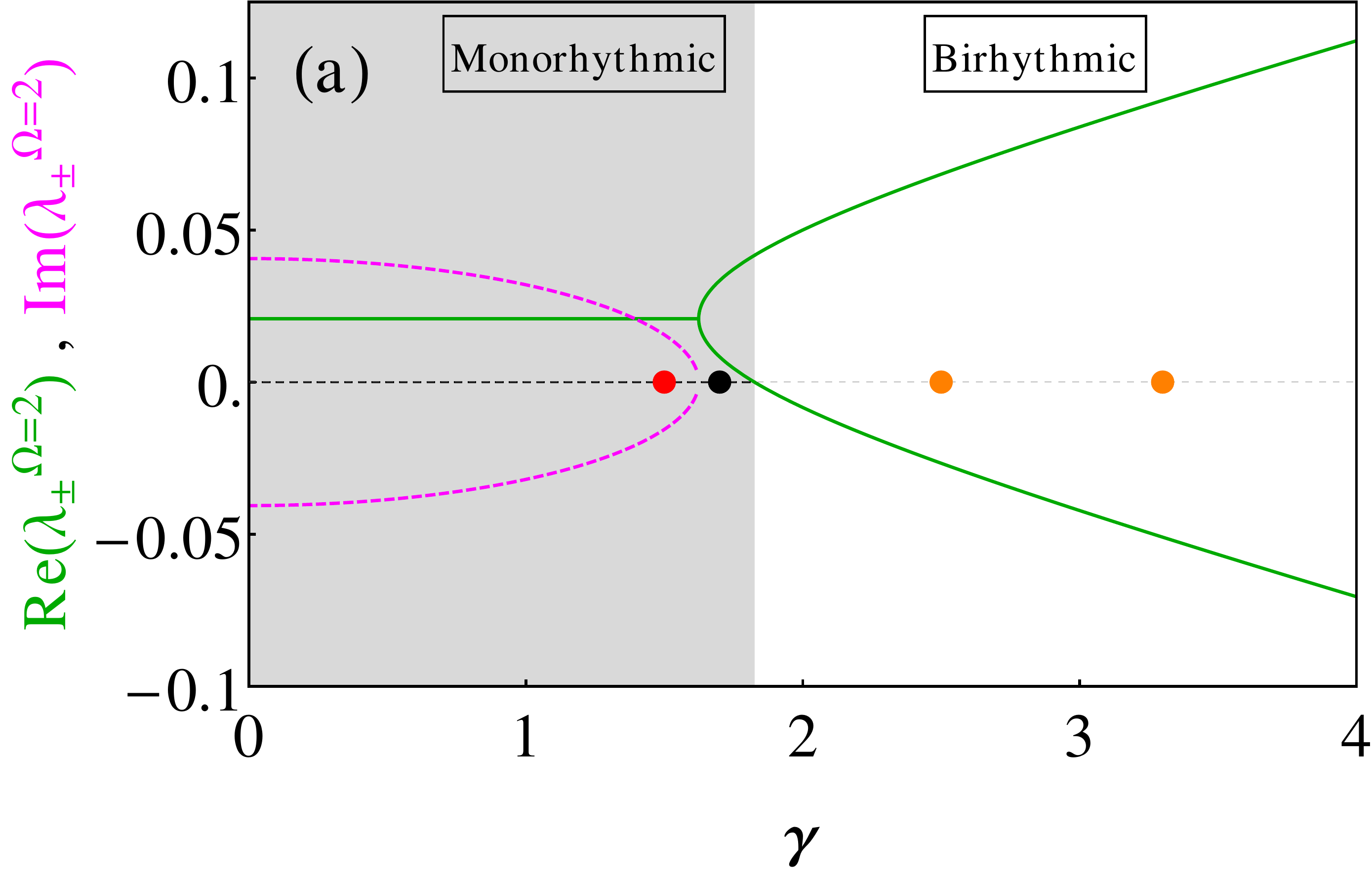}\\
\includegraphics[width=4cm, height=4cm]{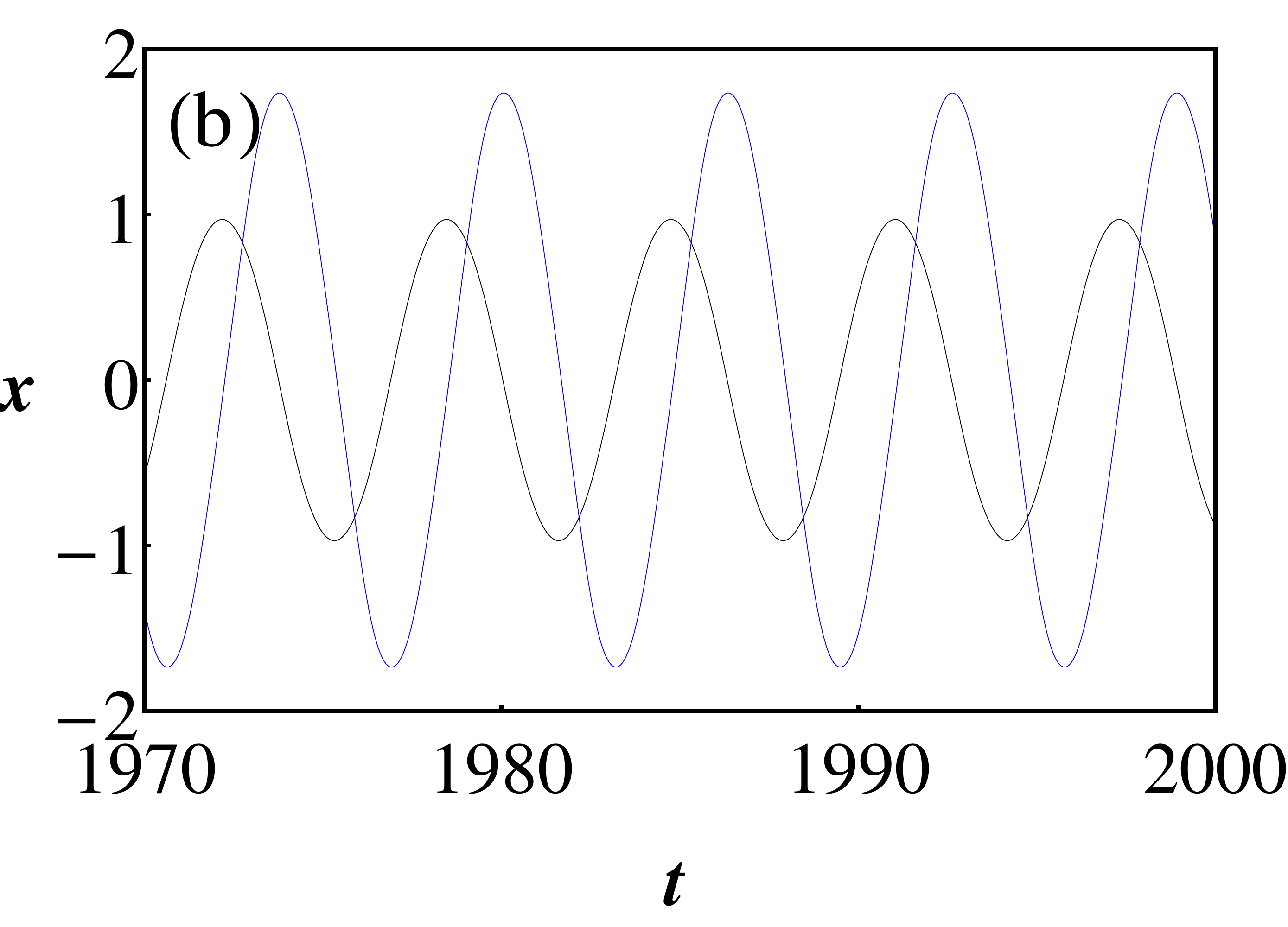}
\includegraphics[width=4cm, height=4cm]{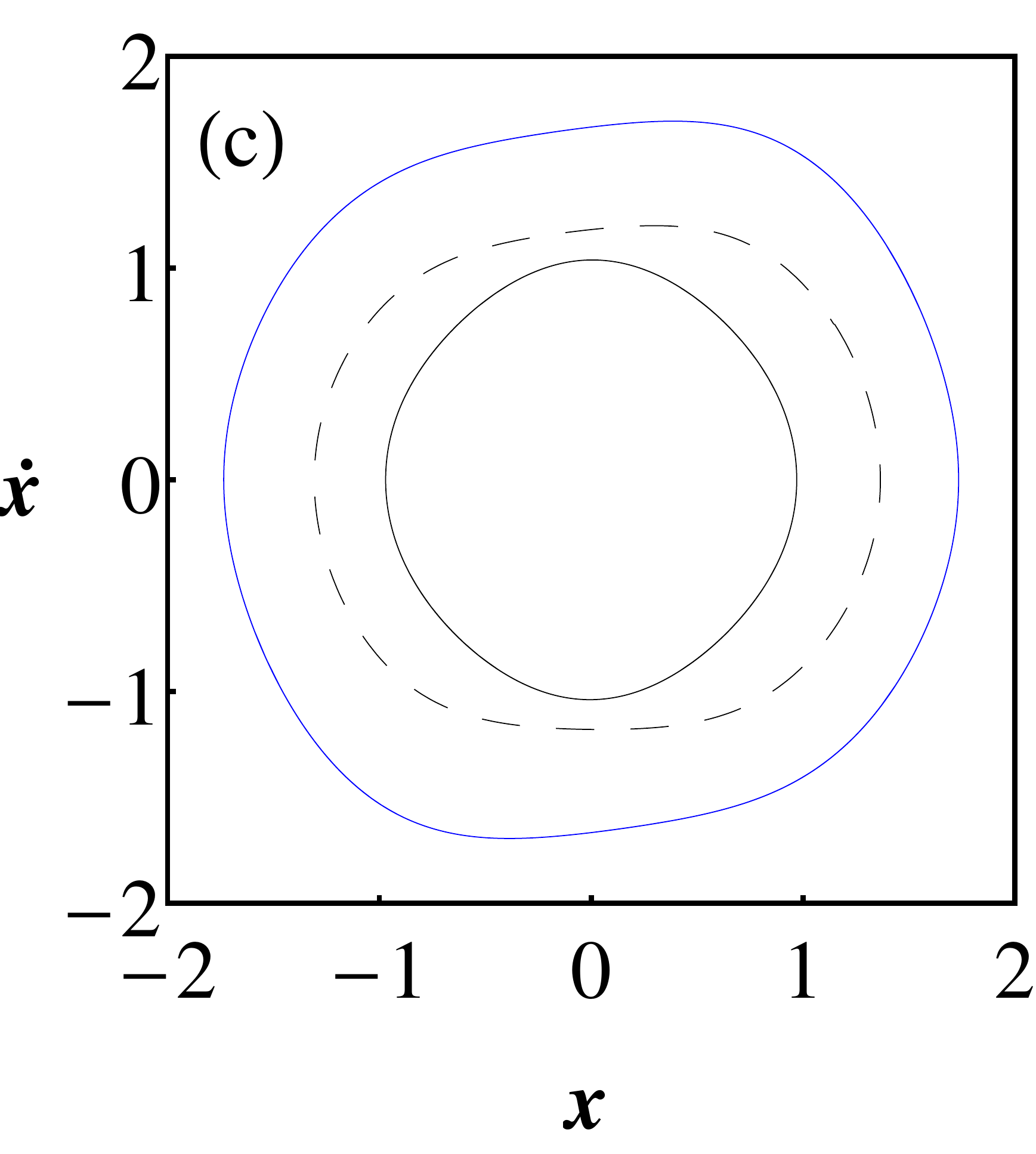}
\caption{ \SC{\emph{Birhythmic response of PENVO with delay.} Sublot (a) exhibits the birhythmic range of the PENVO with delay with the variation of the parameter $\gamma$; Eq.~(\ref{eq:pentdvo}) gives monorhythmic behaviour in the (gray shaded) range: $\gamma \in[0,1.82]$ (approximately), beyond which the birhythmicity starts. The corresponding eigenvalues---found after linear stability analysis about the origin in the $p$-$q$ plane---are complex conjugate pairs with positive real part (green solid line) for $0 \le \gamma < 1.62$ (approximately) and purely real positive numbers for $1.62 \le \gamma \le 1.82$ (approximately). (The imaginary parts of the eigenvalues are depicted using magenta dashed line.) Thereafter the system becomes birhythmic as the pair of eigenvalues become purely real with alternate signs. This subplot should be seen in conjunction with Fig.~\ref{fig:stabilityswitching}; the dots making the origin there have been put here on the horizontal gray dashed line for the convenience of comparison. Furthermore, the time series plot (b) and the  phase space plot (c) for Eq.~(\ref{eq:pentdvo}) with $\gamma=3.3$ explicitly illustrate the existence of the birhythmic oscillations. The {blue solid} {{and the {black solid}}} lines correspond to two different initial conditions $(1.056,~-0.8576)$ and $(1.576,~1.037)$, respectively. An unstable limit cycle (black dashed line) is also present between the two stable limit cycles. The other parameter values used in the figure are $K=\mu=0.1,\,\tau=0.623$, and  $\Omega=2$.}}
\label{fig:birhythmic} 
\end{figure}

Now, we ask the question if the system allows for birhythmicity. We realize that a convenient way to search for it is to look for stable fixed points ({except the one at the origin}) and stable limit cycles in the corresponding $p$-$q$ plane. A closer look at Eqs.~(\ref{eq:pentdvo_vdp_plane_2}) and (\ref{eq:pentdvo_vdp_plane_4}) reveals that $(0,0)$ is a common fixed point and, additionally, we have seen that they possess limit cycles. Straightforward linear stability analysis about the fixed point for the case $\Omega=4$ yields $\left(\mu \pm{i K e^{\pm i \tau } }\right)/2$ as the eigenvalues that clearly has real negative part and there is no local bifurcation possible with change in $\gamma$. In fact, detailed numerical study suggests that, for the appropriately fixed parameters and $\Omega = 4$, no changes occur except that the oscillation in the amplitude of the limit cycle becomes less perceptible with increase in $\gamma$. Naturally, one expects only monorhythmicity in the system. 

The case of $\Omega=2$ is, however, very interesting: The linear stability about $(0,0)$ yields the eigenvalues, $\SC{\lambda_{\pm}^{\Omega=2}=}(\pm \sqrt{\gamma ^2 \mu ^2-2 K^2 \cos (2 \tau)-2 K^2}-2 K \sin \tau +2 \mu)/4$,   and thus the character of the fixed point can change with the value of $\gamma$, e.g., it is quite clear that for small values of $\gamma$ (other parameters being appropriately fixed) the origin should be a focus and for larger values it should be a saddle. The full study of Eq.~(\ref{eq:pentdvo_vdp_plane_2}) being analytically quite cumbersome, we present a numerical illustration of how birhythmicity is generated by varying $\gamma$.

\begin{figure}
\includegraphics[width=8cm, height=6cm]{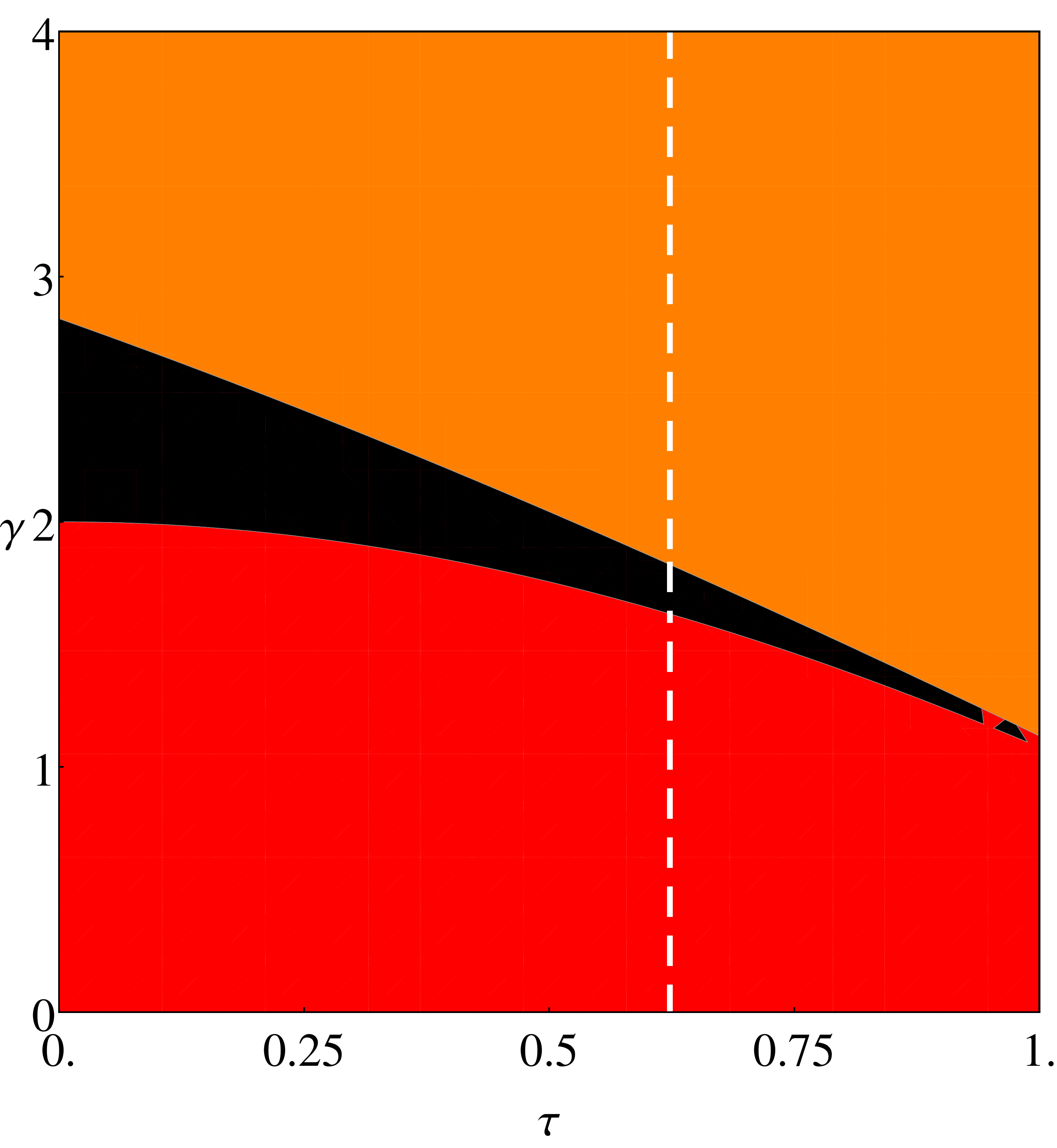}
\caption{\SC{\emph{Interplay of delay and excitation strength, $\gamma$, for PENVO with delay.}
Plot depicts how the birhythmic region ({orange}) increases with delay. The red and the black regions correspond to the monorhythmic states where the origin in the corresponding $p$-$q$ plane is unstable focus and unstable node respectively. The vertical white line marks the transition from the monorhythmic to birhythmic oscillations as studied in  Fig.~\ref{fig:stabilityswitching} and Fig.~\ref{fig:birhythmic} for $\tau=0.623$. The values of the relevant parameters used in this figure are $\mu=0.1$ and $K=0.1$. }}
\label{fig:birhythmic_zone_Tau_Gamma} 
\end{figure}

In this respect, please see Fig.~\ref{fig:stabilityswitching} \SC{(and also Fig.~\ref{fig:birhythmic}(a))} where we have depicted the vector plots corresponding to Eq.~(\ref{eq:pentdvo_vdp_plane_2}) for $\gamma=1.5$, \SC{$\gamma=1.7$}, $\gamma=2.5$, and $\gamma=3.3$. We have fixed $\Omega=2$, $\tau=0.623$, and $K=\mu=0.1$. \SC{Careful study reveals that, as $\gamma$ is increased, after $\gamma \approx1.62$ the origin becomes an unstable node from an unstable focus; the limit cycle still exists. On further increasing $\gamma$ to approximately $1.82$, the origin becomes a saddle from the unstable node.} The saddle however is born along with two stable foci (say, $F_1^-$ and $F_1^+$) at which the stable manifolds of the saddle terminate; two other stable foci are also born (say, $F_2^-$ and $F_2^+$) and the limit cycle, that exists around the origin for {$\gamma\lesssim1.82$}, is annihilated.  One observes that at a given $\gamma$, the value of $p^2+q^2$ is same for $F_1^-$ and $F_1^+$, and also for $F_2^-$ and $F_2^+$, meaning that only two (and not four) different limit cycles can be observed in the PENVO with delay when $\gamma \gtrsim1.82$. We verify this conclusion by numerically solving Eq.~(\ref{eq:pentdvo}) for two different initial conditions but {at the same set of parameter values} and as shown in Fig.~\ref{fig:birhythmic}, we observe birhythmic oscillations. To conclude what we have shown is that by changing $\gamma$ we can induce birhythmicity or conversely, one can say that if the system is already birhythmic, \emph{we can make the system monorhythmic by using $\gamma$ as a control parameter.} \SC{It is interesting to note that on varying both $\gamma$ and $\tau$, the states of birhythmic oscillations appear over a wider region in $\tau$-$\gamma$ space as shown in Fig.~\ref{fig:birhythmic_zone_Tau_Gamma}. For all values of delay the aforementioned mechanism behind appearance of birhythmic states with change in $\gamma$ is same: the corresponding limit cycle attractor around the origin in the $p$-$q$ plane makes way for four fixed point attractors.}

\section{Multicycle PENVO} \label{KBM}
Up to now we have seen how a delay term added in the PENVO modifies the antiresonance and the resonance at $\Omega=2$ and $\Omega=4$ respectively, and furthermore, gives rise to birhythmicity that in turn can be controlled by the strength of the periodically modulated nonlinearity in PENVO. Another natural modification of the van der Pol oscillator with multiple limit cycles is a variant of the van der Pol oscillator---originally proposed~\cite{kaiser83,kaiser91} to model enzyme reaction in biochemical system---with a {sextic} order polynomial as damping coefficient:
\begin{eqnarray}
\ddot{x}+\mu (-1+x^2-\alpha x^4+\beta x^6) \dot{x}+x=0.
\label{eq:kaiser}
\end{eqnarray}
Here, $  0<\mu \ll 1$ and $\alpha, \beta>0$. We call it Kaiser oscillator. It has three concentric limit cycles surrounding an unstable focus at the origin:  two of them are stable and the unstable one acts as the boundary separating the basins of attractions of the two stable cycles. However, whether there are two stable limit cycles (birhythmicity) or only one (monorhythmicity) strictly depends on values of $\alpha$ and $\beta$. Under the assumption that $\mu\ll1$, straightforward application of the Krylov--Bogoliubov method helps to demarcate the regions of birhythmicity and monorhythmicity in $\alpha-\beta$ parameter space (see Fig.~\ref{fig:birhythmicity_switch} in Appendix~\ref{sec:Kaiser-parameter}).
\begin{figure}
\includegraphics[width=4.15cm, height=3.7cm]{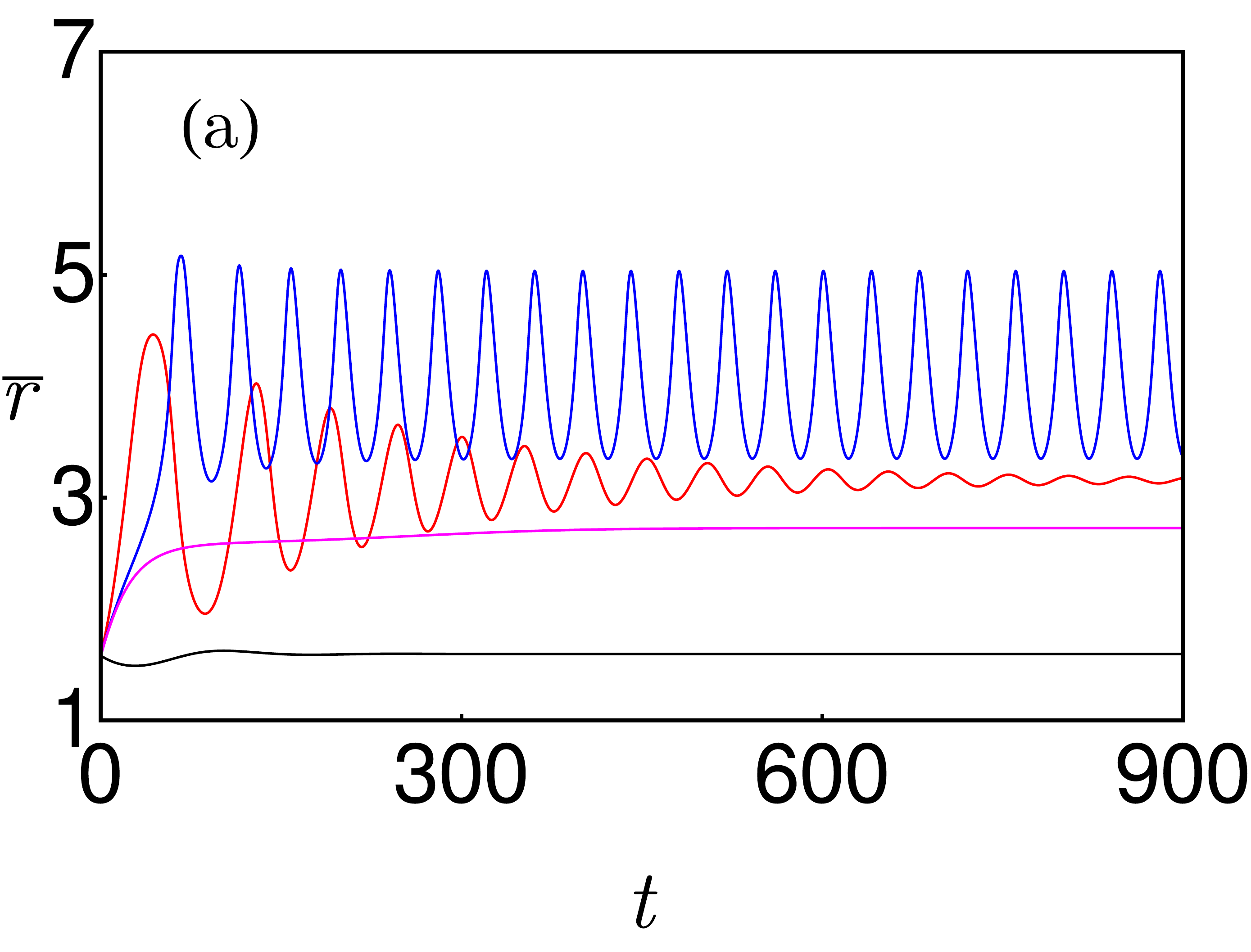}~
\includegraphics[width=4.15cm, height=3.7cm]{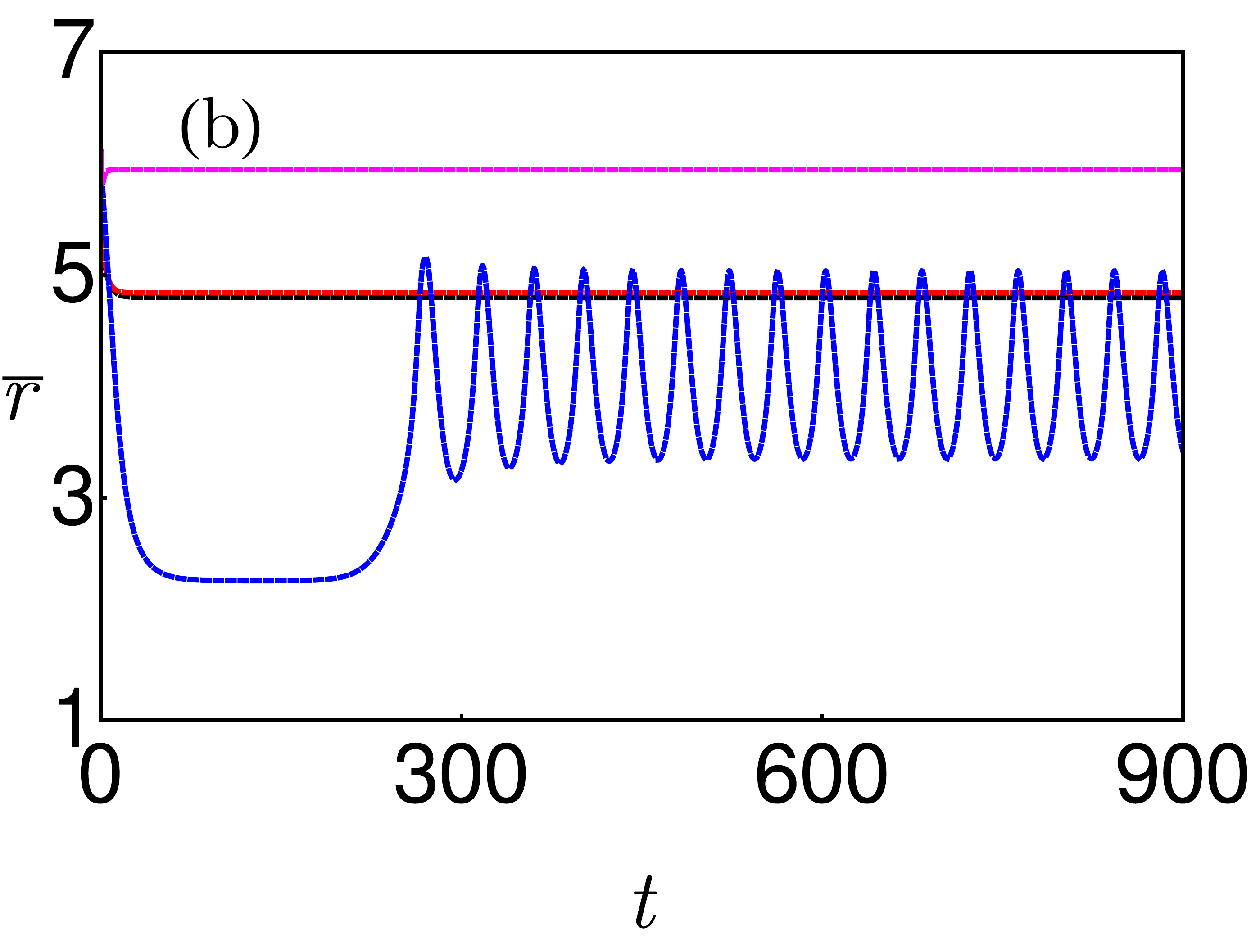}\\
\hspace{-9 pt}\includegraphics[width=4.3cm, height=3.4cm]{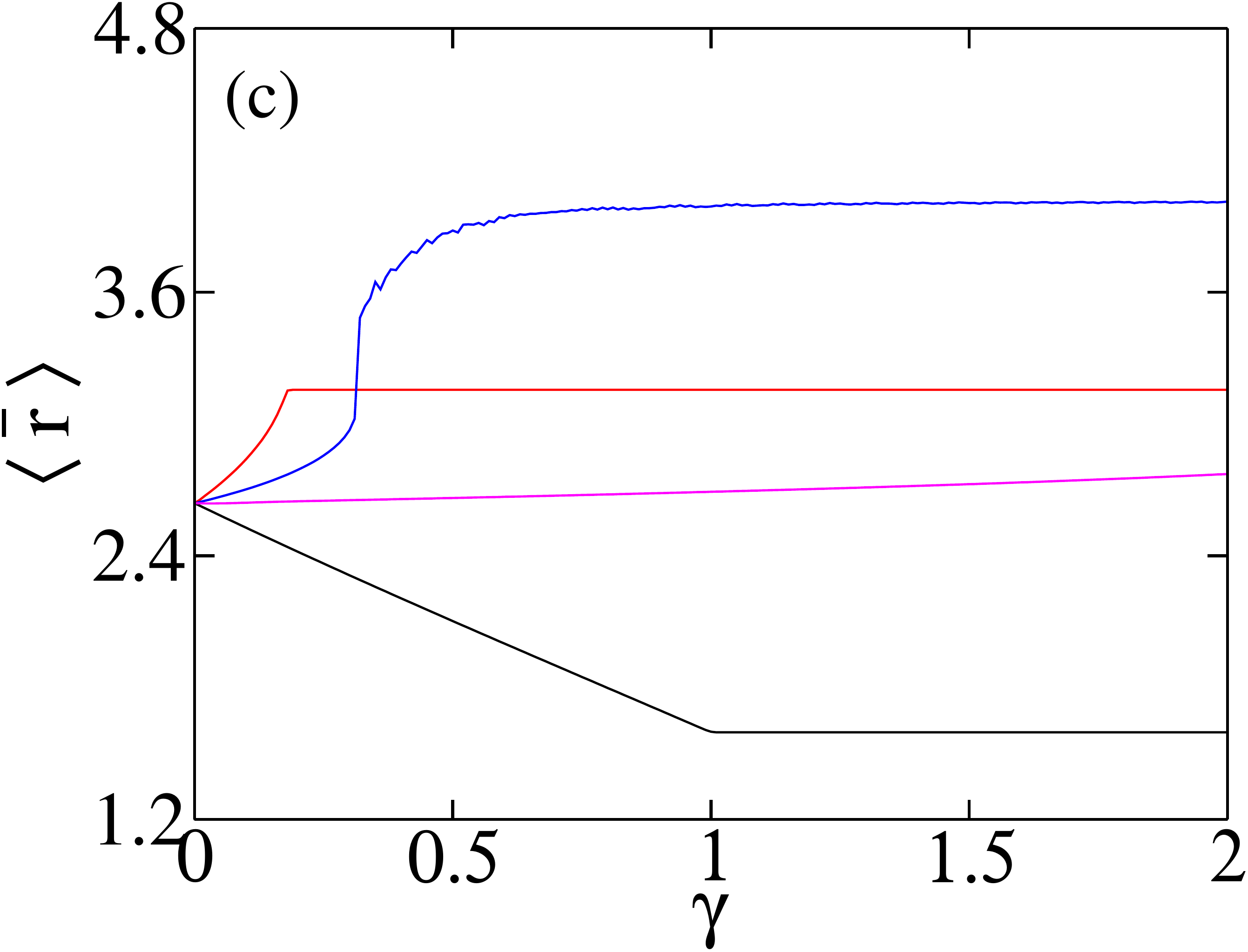}~
\includegraphics[width=4.25cm, height=3.35cm]{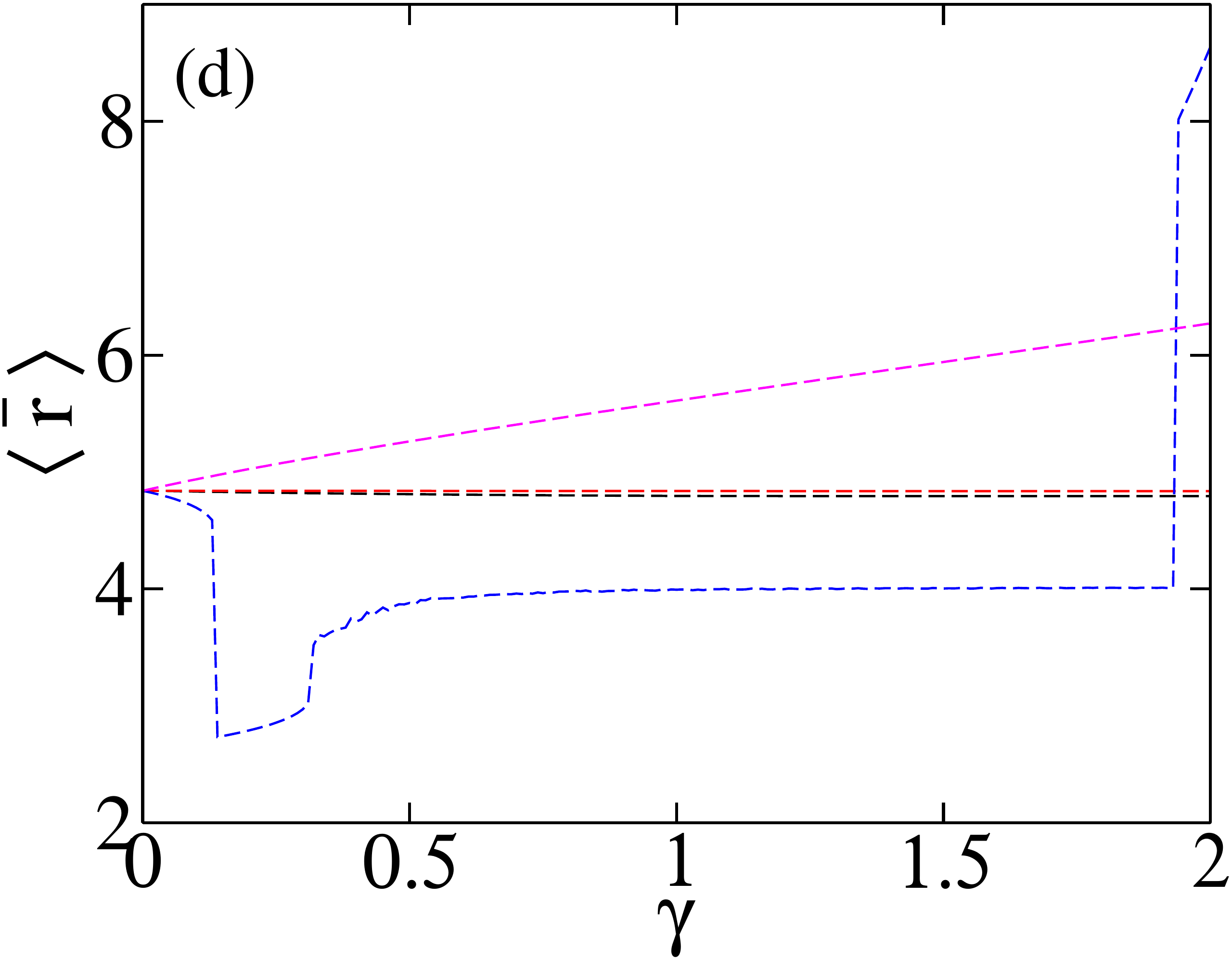}
\caption{ \emph{Resonant and antiresonant responses in multicycle PENVO.} Presented are time series plots (subplot a and b) corresponding to both small (solid line) and large (dotted line) cycles for $\Omega=2~{\rm (black)},~4~{\rm(red)},~6~{\rm (blue)~and}~8~{\rm(magenta)}$. Furthermore, subplots (c) and (d) depict how the averaged amplitudes of the responses change with $\gamma\in[0,2]$. It is depicted that the smaller limit cycle shows resonances for the case $\Omega=4,\,6$ and $8$  but antiresonance for the case $\Omega=2$; the larger limit cycle admits resonance for $\Omega=8$ but antiresonance for the case $\Omega=2,\,4$ and $6$. The values of the parameters used to numerically solve Eq.~(\ref{eq:kaiser_pe_amp_ph}) for the purpose of the figure are $\alpha=0.144,~\beta=0.005,~\mu=0.1~\text{and}~\gamma=1.5$ (in subplot a and b).}
\label{fig:kaiser_pe_amp_vs_gamma} 
\end{figure}
In the context of this paper, it is of immediate curiosity to ponder upon the important questions like `can one find resonance and antiresonance in the Kaiser oscillator', `would periodically modulating the nonlinearity control the inherent birhythmicity in the Kaiser oscillator', etc. 

The addition of the periodic modulation of nonlinearity in the Kaiser oscillator get us the following equation:
\begin{eqnarray}
\ddot{x} + \mu \left[1+\gamma \cos(\Omega t)\right] (-1+x^2-\alpha x^4+\beta x^6) \dot{x} + x &=0, \,\quad\label{eq:kaiser_pe}
\end{eqnarray}
where $\gamma> 0$. For obvious reasons, henceforth we aptly call this system: multicycle PENVO. Again, the Krylov--Bogoliubov method yields,
\begin{subequations}
\begin{eqnarray}
\dot{\overline{r}} &=& \frac{1}{128} \overline{r} \mu  \left(-5 \beta  \overline{r}^6+8 \alpha  \overline{r}^4-16 \overline{r}^2+64\right) +A_\Omega(\overline{r},\overline{\phi};\gamma),\,\,\,\quad\\
\dot{\overline{\phi}} &=& B_\Omega(\overline{r},\overline{\phi};\gamma)+O(\mu^2).
\end{eqnarray}
\label{eq:kaiser_pe_amp_ph}
\end{subequations}
Here the symbols are in their usual meaning as detailed in Sec.~\ref{sec2}. The subscripts specify the value of $\Omega$ at which $A_\Omega$ and $B_\Omega$ have to be determined; the functions have singularities at $\Omega = 2,\,4,\,6$ and $8$, and their limiting values at these $\Omega$-values are respectively,
\begin{subequations}
\begin{eqnarray}
 A_2 &=& -\frac{1}{64} \gamma  \overline{r} \mu  \cos (2 \overline{\phi} ) \left(\beta  \overline{r}^6-\alpha  \overline{r}^4+16\right),
 \\
 B_2 &=& -\frac{1}{64} \gamma  \mu  \sin (\overline{\phi} ) \cos (\overline{\phi} ) \left(7 \beta  \overline{r}^6-10 \alpha  \overline{r}^4+16 \overline{r}^2-32\right); \nonumber\\ \\
 A_4 &=&\frac{1}{64} \gamma  \overline{r}^3 \mu  \cos (4 \overline{\phi} ) \left(\beta  \overline{r}^4-2 \alpha  \overline{r}^2+4\right), 
 \\
  B_4& =& -\frac{1}{128} \gamma  \overline{r}^2 \mu  \sin (4 \overline{\phi} ) \left(7 \beta  \overline{r}^4-8 \alpha  \overline{r}^2+8\right); \\
A_6 &= &-\frac{1}{64} \gamma  \overline{r}^5 \mu  \cos (6 \overline{\phi} ) \left(\alpha -\beta  \overline{r}^2\right), \\
B_6 &= &\frac{1}{128} \gamma  \overline{r}^4 \mu  \sin (6 \overline{\phi} ) \left(2 \alpha -3 \beta  \overline{r}^2\right);\\
A_8 &=&\frac{1}{256} \beta  \gamma  \overline{r}^7 \mu  \cos (8 \overline{\phi} ), \\
B_8 &= &-\frac{1}{256} \beta  \gamma  \overline{r}^6 \mu  \sin (8 \overline{\phi} ).
\end{eqnarray}
\label{eq:kaiser_pe_amp_ph_gamma}
\end{subequations}
\begin{figure*}
\includegraphics[width=4.1cm, height=4.1cm]{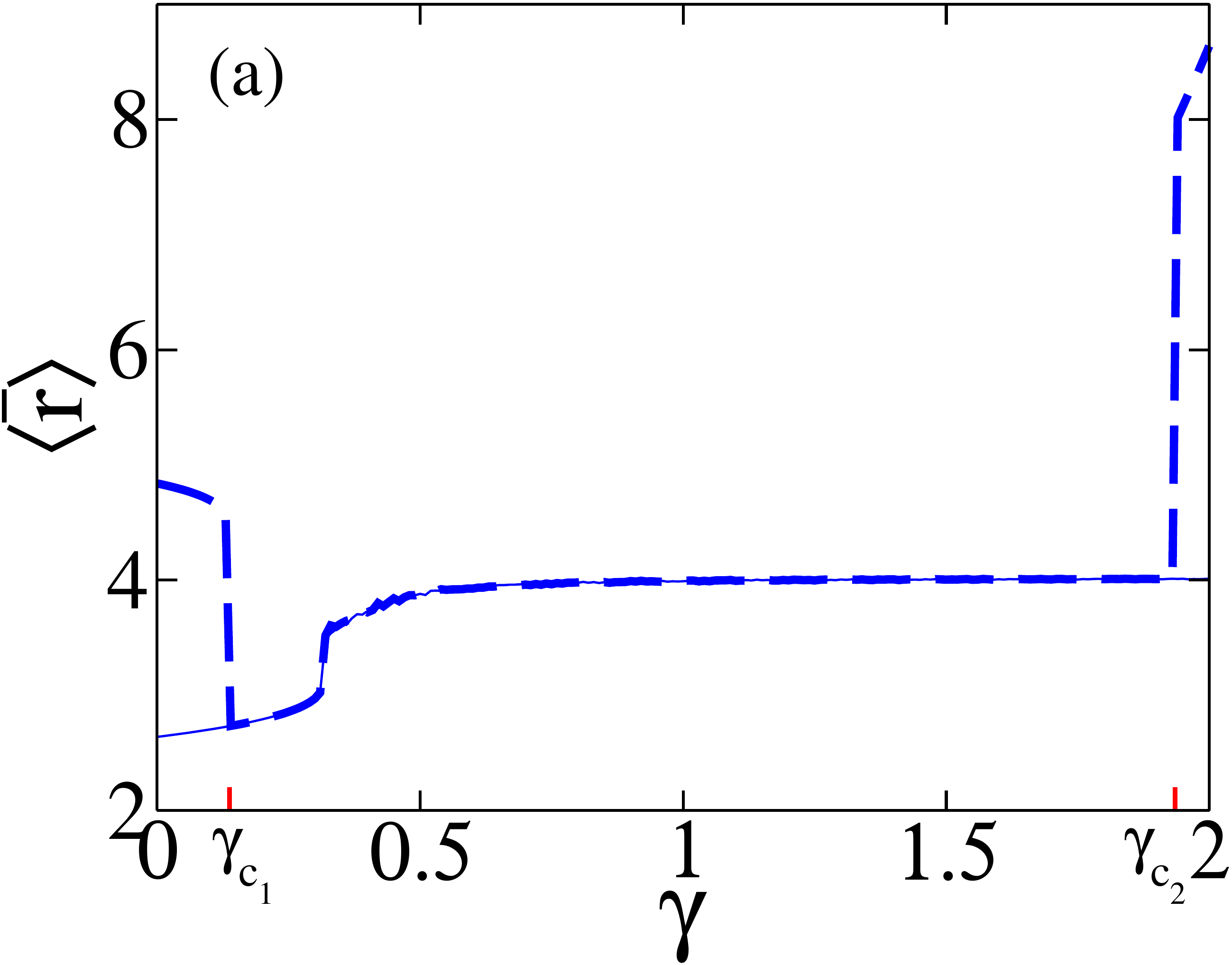}
\includegraphics[width=4.1cm, height=4.1cm]{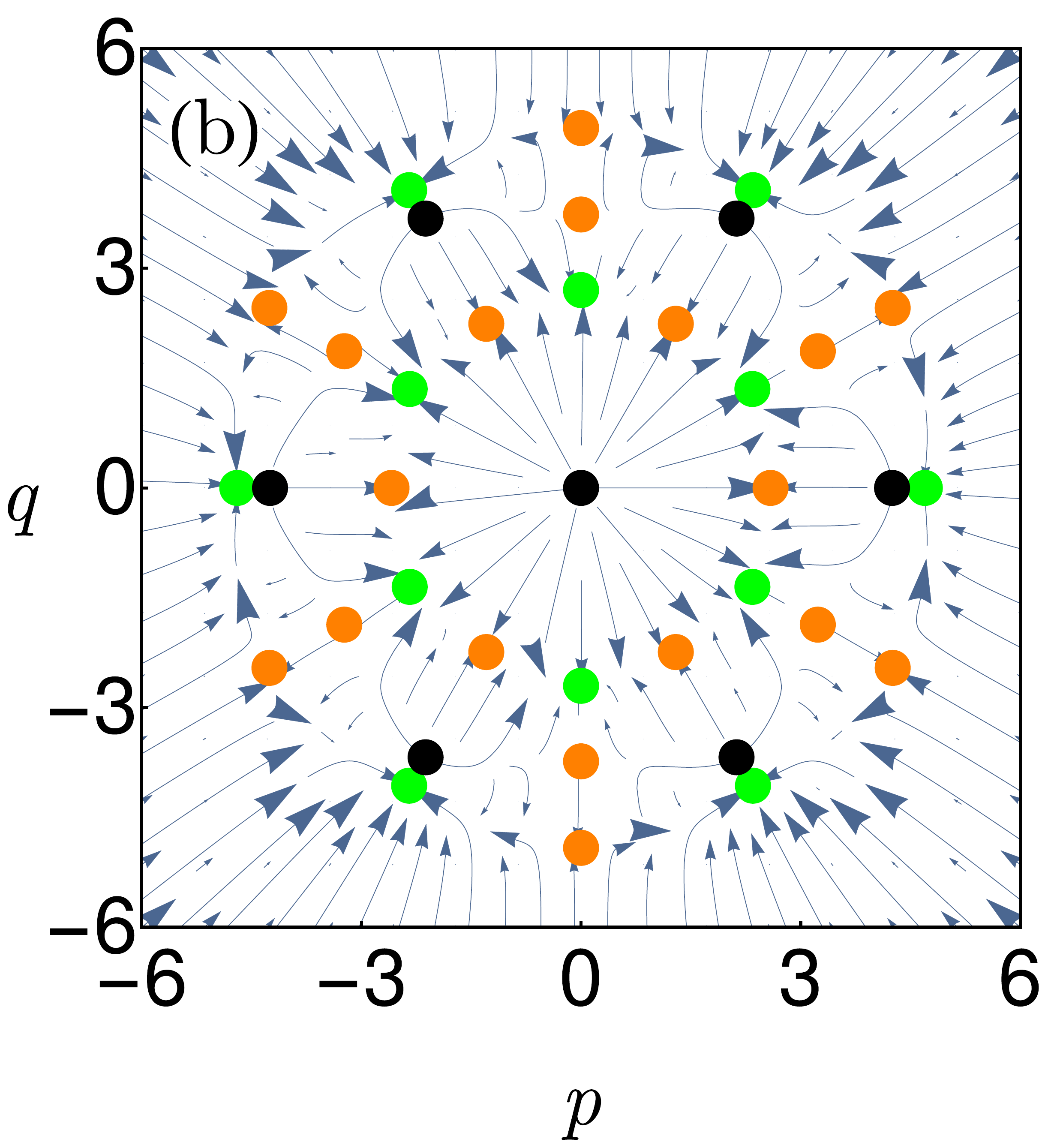}
\includegraphics[width=4.1cm, height=4.1cm]{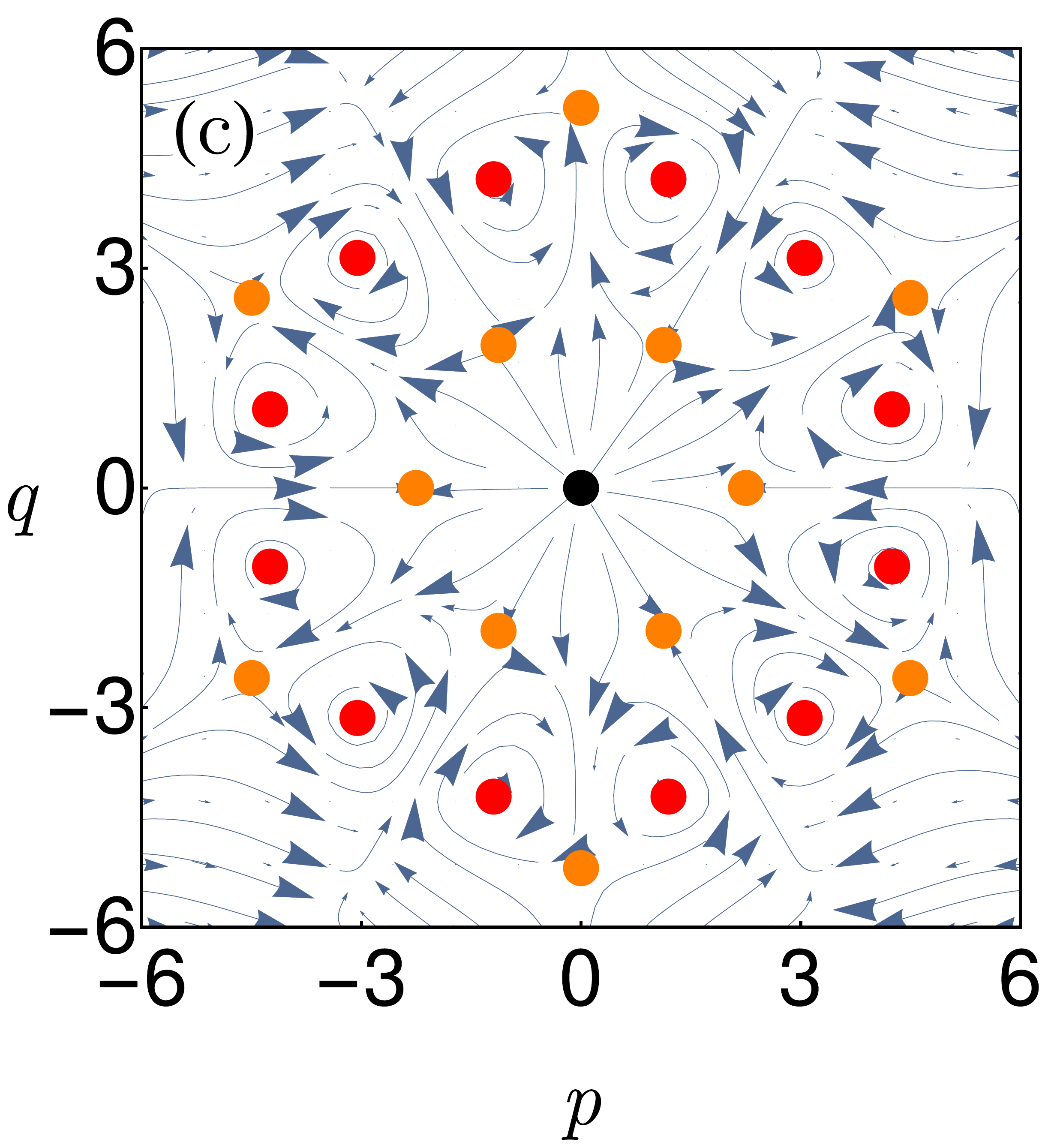}
\includegraphics[width=4.1cm, height=4.1cm]{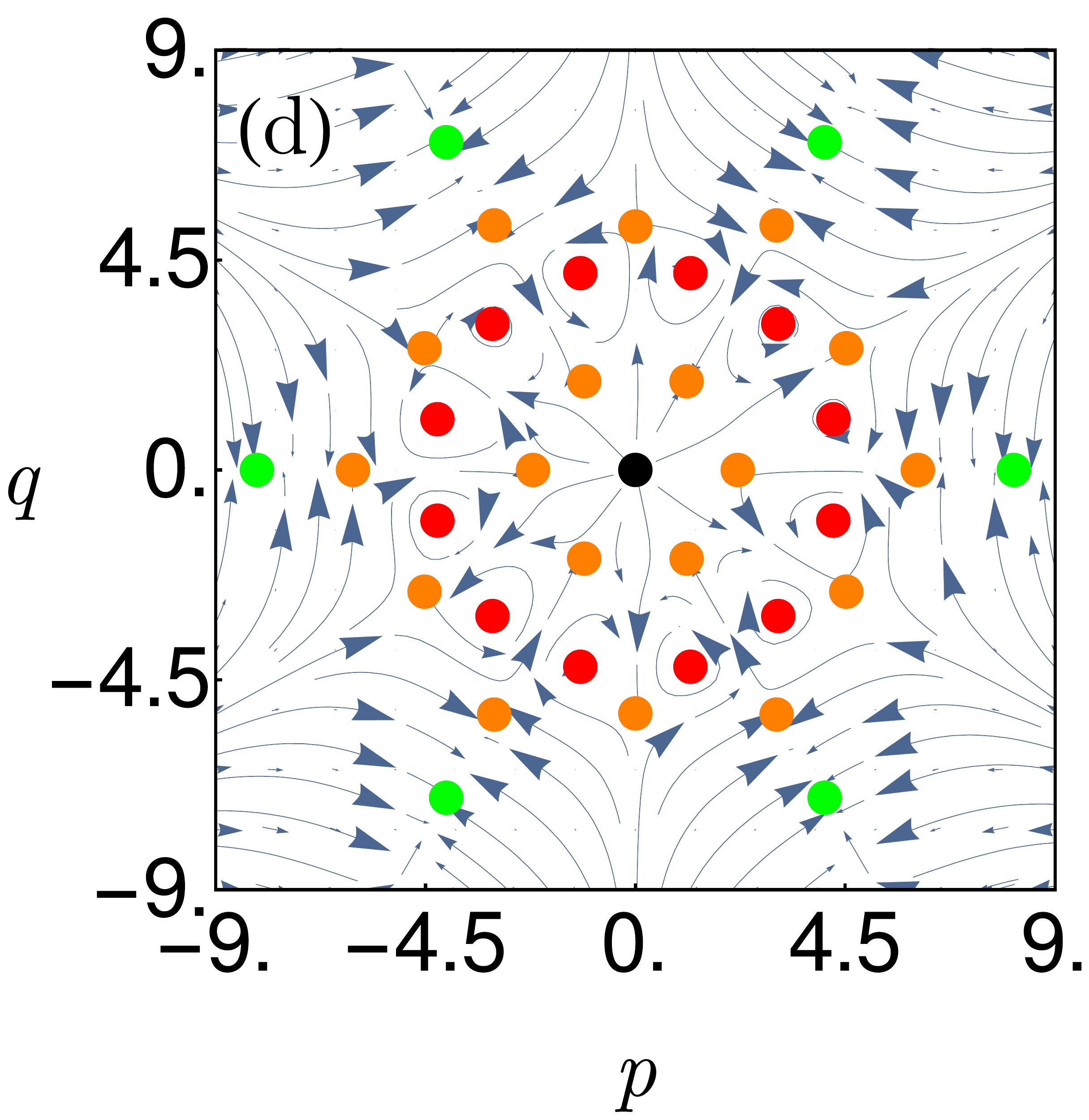}\caption{\emph{Strength of periodic modulation of nonlinear damping controls birhythmicity in multicycle PENVO.} Subplot (a) presents the observation that the average amplitudes of the periodic responses---the smaller limit cycle (solid blue line) and the larger limit cycle (dotted blue line)---merge for an intermediate range of $\gamma$ between $\gamma_{c_1}\approx0.138$  to $\gamma_{c_2}\approx1.935$ resulting in monorhythmicity. Streamplots (b)-(d) depict repellers [unstable node (black dot) , unstable focus (red dot) and saddle (orange dot)] and attractors [stable node (green dot) and stable limit cycle (around each red dot; not explicitly shown)] in $p$-$q$ space of the multicycle PENVO at $\gamma={0.1,~1.5,~\textrm{and}~1.95,}$ respectively. Other parameter values have been fixed at $\alpha=0.144,~\beta=0.005,~\mu=0.1$ and $\Omega=6$. In subplot (b), there are two sets of stable foci with two distinct values of $\sqrt{p^2+q^2}$ (hence birhythmicity), while in subplot (c) only attractor (and hence monorhythmicity) is a limit cycle---a circle that passes through all the unstable foci with same $\sqrt{p^2+q^2}$-values and centred at origin. In subplot (c), in addition to this limit cycle, another set of stable foci appear with same $\sqrt{p^2+q^2}$-value (hence birhythmicity).}
\label{fig:kaiser_vdP_plane} 
\end{figure*}

As before, we go on to $p$-$q$ plane to recast set of equations~(\ref{eq:kaiser_pe_amp_ph}) for all four $\Omega$-values in terms of $p$ and $q$ variables (see Appendix~\ref{sec:floweqns}) in order to understand the dynamics conveniently. For all the four values of $\Omega$, the origin---$p,q$=(0,0)---is a fixed point that on doing linear stability analysis, turns out to be unstable for all values of $\gamma$. Since now the corresponding equations of motion are much more cumbersome to handle analytically, we resort to a numerical investigation of the systems. First however we need to pick appropriate value of $\alpha$ and $\beta$. We choose $\alpha=0.144$ and $\beta=0.005$ that would allow the Kaiser oscillator (multicycle PENVO with $\gamma=0$) to exhibit birhythmicity (see Appendix~\ref{sec:Kaiser-parameter}); the amplitudes of the limit cycles that are concentric circles about $(x,\dot{x})=(0,0)$ in the limit $\mu\rightarrow0$ are approximately $2.64$ and $4.84$ respectively. In what follows, we work with $\mu=0.1$.

We now turn on the periodic modulation of the nonlinear term, i.e., we work with the multicycle PENVO with nonzero $\gamma$. We scan the system for various values of $\gamma$ and present the results for $\gamma$ up to $2$ in Fig.~\ref{fig:kaiser_pe_amp_vs_gamma}. For illustrative purpose, consider $\gamma=1.5$. We note that the amplitude of the smaller limit cycle of the Kaiser oscillator increases for the case $\Omega=4,\,6$ and $8$ (resonances) but decreases for the case $\Omega=2$ (antiresonance). Similarly, while the amplitude of the larger limit cycle of the Kaiser oscillator increases for the case $\Omega=8$ (resonance), but it decreases for the case $\Omega=2,\,4$ and $6$ (antiresonances).  As an aside, for the case $\Omega=6$, we also note that the amplitudes of both the cycles themselves oscillate and the response corresponding to the outer limit cycle changes from antiresonance to resonance as $\gamma$ increases (see Fig.~\ref{fig:kaiser_pe_amp_vs_gamma}d).

More interesting, however, is the fact that the resonance and the antiresonance, manifested as limit cycles with oscillating amplitudes, for $\Omega=6$ merge---as implicitly shown in Fig.~\ref{fig:kaiser_vdP_plane}(a)---for a range of $\gamma$-values: $\gamma\in(\gamma_{c_1},\gamma_{c_2})\approx(0.138,1.935)$.
This means that \emph{$\gamma$ is yet again acting as a control parameter in bringing about monorhythmicity by suppressing the birhythmicity.} To understand the phase dynamics of control of the aforementioned birhythmicity, we consider the system (\ref{eq:kaiser_pe_amp_ph}) in $(p,~q)$ plane at three representative values of $\gamma$, viz., $\gamma=0.1$~(Fig.~\ref{fig:kaiser_vdP_plane}b), $\gamma=1.5$~(Fig.~\ref{fig:kaiser_vdP_plane}c), and $\gamma=1.95$~(Fig.~\ref{fig:kaiser_vdP_plane}d).  For $\gamma=0.1<\gamma_{c_1}$, a case of birhythmicity, there are twelve stable nodes---the only attractors in the phase space---that can be classified into two groups such that one group of nodes has $\sqrt{p^2+q^2}\approx2.70$ and the other group has $\sqrt{p^2+q^2}\approx4.67$. This corresponds to the fact that there are two distinct limit-cycles in the $x$-$\dot{x}$ plane, and their radii are $2.70$ and $4.67$; in other words, the system is birhythmic. In the monorhythmic case of $\gamma=1.5\in(\gamma_{c_1},\gamma_{c_2})$, we note that the attractors now are twelve limit cycles whose centers (unstable focus) lie on a circle of radius $4.38$ (approximately). Thus, the system has now become monorhythmic and the limit cycle in the $x$-$\dot{x}$ plane has periodically oscillating amplitude. The bifurcation leading to the creation of the twelve symmetrically placed limit cycles takes place at $\gamma=\gamma_{c_1}$ when the stable nodes and the unstable saddles (present at $\gamma<\gamma_{c_1}$) merge appropriately to give rise to the limit cycles (seen at $\gamma>\gamma_{c_1}$). Finally, For $\gamma=1.95>\gamma_{c_2}$, the system showcases {birhythmic} behaviour yet again: the six symmetrically placed asymptotically stable nodes in the corresponding $p$-$q$ plane have identical values for $\sqrt{p^2+q^2}$, \emph{viz.}, $8.12$ that corresponds to the amplitude of the limit cycle of the multicycle PENVO.
\begin{figure}
\includegraphics[width=4cm, height=4cm]{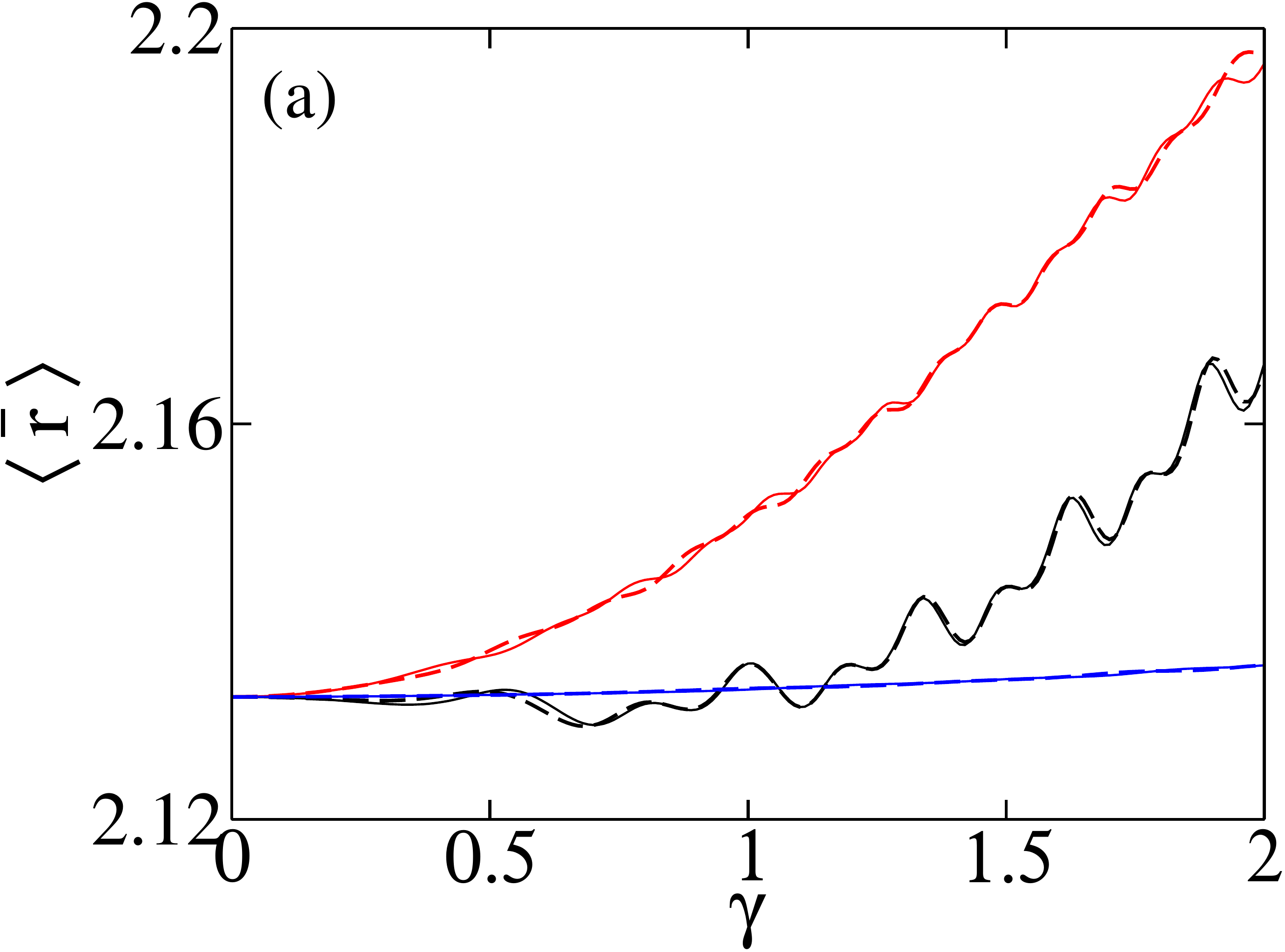}
\includegraphics[width=4cm, height=4cm]{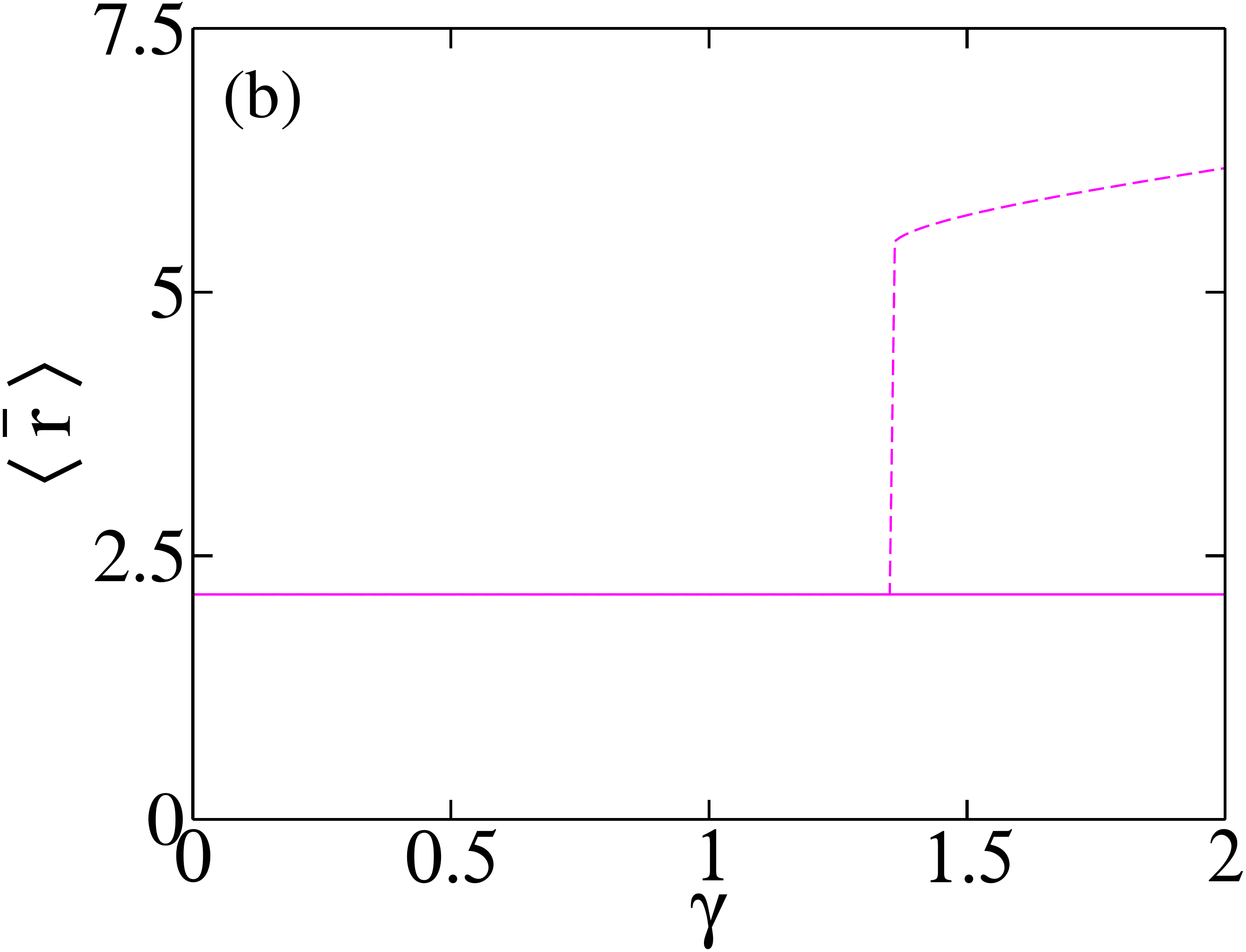}\\~\\
\includegraphics[width=6.5cm, height=4.5cm]{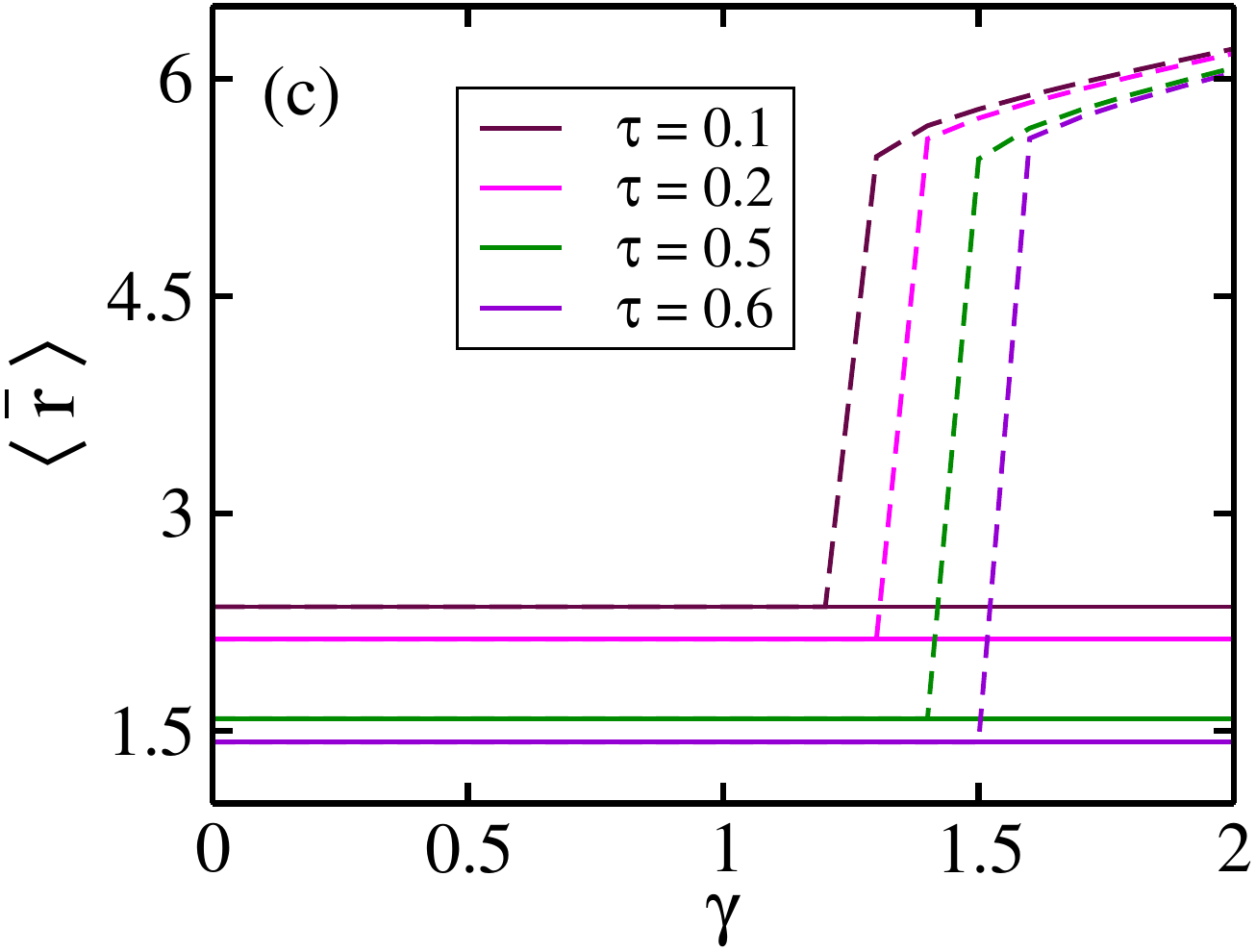}
\caption{\emph{Controlling birhythmicity via delay in multicycle PENVO.}
Subplots (a) and (b) exhibit how the averaged amplitudes change with $\gamma\in[0,2]$ corresponding to both small (solid line) and large (dotted line) cycles for $\Omega=2~{\rm (black)},~4~{\rm(red)},~6~{\rm (blue)~and}~8~{\rm(magenta)}$. The values of the relevant parameters used in the figure are $\alpha=0.144,~\beta=0.005,~\mu=0.1~\text{and}~\tau=0.2$. \SC{Subsequently,  subplot (c) shows the interplay of delay and excitation strength, $\gamma$, as some different values of $\tau$---$\tau=0.1~{\rm (maroon)},~0.2~{\rm(magenta)},~0.5~{\rm (green)~and}~0.6~{\rm(violet)}$---are picked for $\Omega=8$. The plot explains how the average amplitudes change with $\gamma\in[0,2]$ corresponding to both small (solid lines) and large (dotted lines) limit cycles. The average amplitude is lower for higher delay at a given value of $\gamma$}.}
\label{fig:kaiser_delay_pe_amp_vs_gamma}
\end{figure}

We note that the birhythmicity present at other resonance and antiresonance conditions, i.e., for $\Omega=2,\,4,\,\textrm{and}\,8$, could not be controlled to monorhythmicity by the variation in $\gamma$. However, recalling that in Sec.~\ref{sec2} the combination of $\gamma$ and delay could effect control of birhythmicity, one is tempted to add delay term, viz., `$-K x(t-\tau)$' in the left hand side of Eq.~(\ref{eq:kaiser_pe}) with a hope to effect control of birhythmicity for 
$\Omega=2,\,4,\,\textrm{and}\,8$. The introduction on such a delay term in the Kaiser oscillator shifts the region of birhythmicity in the $\alpha$-$\beta$ plane (see Appendix.~\ref{sec:Kaiser-parameter}). In the simultaneous presence of non-zero 
$\gamma$ and $K$, the multicycle PENVO's response at $\Omega=2,\,4,\,6,\,\textrm{and}\,8$ can be analyzed using the Krylov--Bogoliubov method just as has been done in detail for Eq.~(\ref{eq:pentdvo}) and Eq.~(\ref{eq:kaiser_pe}). We omit the repetitive details and rather present the summary of the analyses in Fig.~\ref{fig:kaiser_delay_pe_amp_vs_gamma}(a)-(b). We note that the delay does indeed suppress birhythmicity; and interestingly in the case of $\Omega=8$, \emph{$\gamma$ can be seen to be a control parameter even in the presence of delay.} \SC{Interestingly, with increase in the delay, monorhythmicity to birhythmicity transition is effected at even higher values of $\gamma$ as shown in Fig.~\ref{fig:kaiser_delay_pe_amp_vs_gamma}(c).}

\section{Discussion and Conclusions}
\label{C}
How to control birhythmicity in an oscillator is an interesting question. In this paper we have illustrated that the birhythmicity seen in the delayed van der Pol oscillator and the van der Pol oscillator modified to have higher order nonlinear damping (the Kaiser oscillator) can be suppressed if the nonlinear terms of the oscillators are periodically modulated. This periodic modulation of the nonlinear damping also brings about resonance and antiresonance responses in the aforementioned oscillators. In order to characterize the responses, we have presented perturbative calculations using the Krylov--Bogoliubov method and supplemented them with ample numerical solutions for the systems of ordinary differential equations under consideration. We have also discussed in detail how to understand the bifurcations leading to monorhythmicity from birhythmicity (and  vice versa) from the relevant phase space trajectories obtained via the perturbative technique.

We recall that the introduction of delay is one of the popularly known method of controlling birhythmicity. However, as we have seen in Sec.~\ref{sec2}, delay can introduce birhythmicity as well. It is interesting to realize in such cases periodically modifying the nonlinear terms can change the birhythmic behaviour to monorhythmic. A comparison of responses due to delay and parametric excitation in a limit cycle system provides an extra tool-kit for controlling birhythmicity when one alone may not be fruitful. We may point out that the delay term we have used in this paper is completely position dependent as opposed to the more commonly investigated velocity dependent delay terms~\cite{k-dsr,biswas_pre_2016,biswas_chaos_2017} in the literature. 

We strongly believe that the proposed idea of controlling multirhythmicity by invoking periodic modulation of nonlinear terms could be useful in plethora of limit cycle systems. It is also worth pondering if such a mechanism of suppressing multirhythmicity is present in nature because, after all, there is no dearth of the limit cycle oscillations~\cite{Jenkins-2013-PhysicsReports} in nature. However, we do not believe that building a general universal mechanism behind this phenomenon can be proposed easily; each system has to be analysed on a case-by-case basis.  \SC{Also, an interesting future direction of study is to investigate how other controlling schemes such as conjugate self-feedback~\cite{biswas_pre_2016}, self-feedback~\cite{biswas_chaos_2017}, and filtered feedback~\cite{biswas_pre_2019} are affected in the presence of periodic modulation of the nonlinearity in the corresponding systems.}

\section*{Acknowledgment}

SS acknowledges RGNF, UGC, India for the partial financial support. SS is grateful to Rohitashwa Chattopadhyay for his enormous support during a visit to IIT Kanpur, and Pratik Tarafdar for some help with Mathematica. SC is thankful to Anindya Chatterjee (IIT Kanpur) for insightful discussions.

\appendix
{
\section{Birhythmicity in the Kaiser Oscillator: Effect of Delay}
\label{sec:Kaiser-parameter}
Consider the Kaiser model in presence of a position dependent delay: 
\begin{eqnarray}
\ddot{x}+\mu \left( -1+x^2-\alpha x^4+\beta x^6   \right) \dot{x}+ x- K x(t-\tau)=0,\qquad
\label{eq:kaiser_delay}
\end{eqnarray}
$(0 < \epsilon \ll 1;~ \SC{0<\tau<1}$). When $K=0$, the system is either monorhythmic or birhythmic depending on the values of $\alpha$ and $\beta$ as depicted in Fig.~(\ref{fig:birhythmicity_switch}).  It is expected that for small values of $K$ and $\tau$, the behaviour of the Kaiser oscillator should be qualitatively similar, although the region in the $\alpha$-$\beta$ plane where the birhythmic behaviour is seen would be shifted slightly. This is shown in Fig.~(\ref{fig:birhythmicity_switch}) that has been obtained by employing the Krylov--Bogoliubov method to write the equations for the amplitude as well as the phase of the system's response as
\begin{subequations}
\begin{eqnarray}
\dot{\overline{r}} &=& -\frac{\overline{r} \left(64 K \sin \tau+\mu  \left(5 \beta  \overline{r}^6-8 \alpha  \overline{r}^4+16 \overline{r}^2-64\right)\right)}{128},\qquad\quad\\
\dot{\overline{\phi}} &=& -\frac{1}{2} K \cos \tau,
\end{eqnarray}
\label{eq:kaiser_delay_amp_ph}
\end{subequations}
respectively. Here higher order terms have been neglected. It is clear from the existence of non-overlapping regions of birhythmicity that introducing delay may induce monorhythmicity in birhythmic cases or vice versa.
\begin{figure}[h]
\includegraphics*[width=0.3\textwidth]{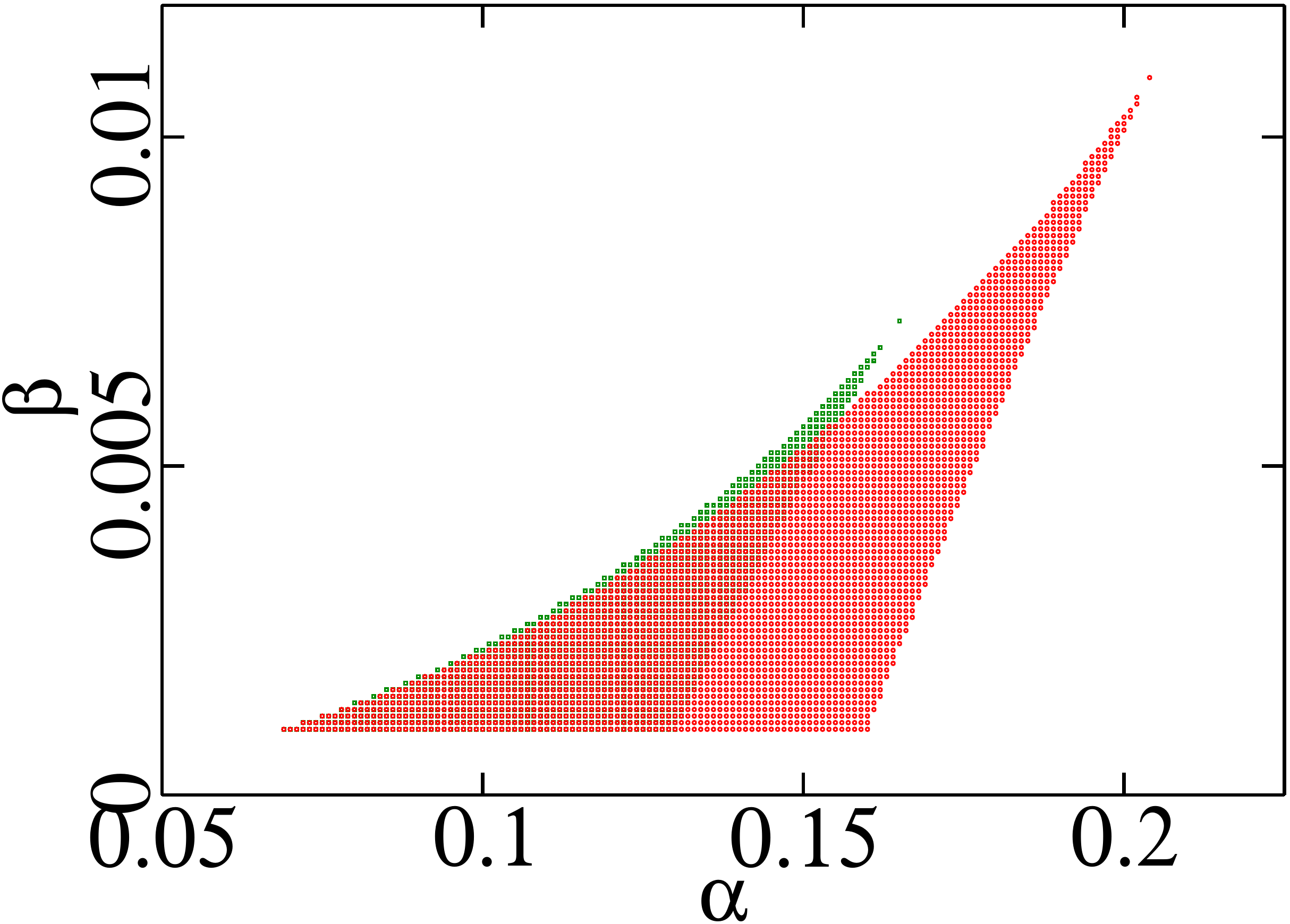}
\caption{\emph{Delay changes rhythmicity.} This figure showcases for what values of $\alpha$ and $\beta$, systems (\ref{eq:kaiser}) and (\ref{eq:kaiser_delay}) are birhythmic---the green and the red zones respectively. In other words, the changes in the birhythmic zone in $\alpha$-$\beta$ parameter space in the presence of the time delay ($K=0.1$ and $\tau=0.2$) have been exhibited. {The systems are monorhythmic when not birhythmic}. Here, $\mu=0.1$.}
\label{fig:birhythmicity_switch}
\end{figure}
\section{Flow Equations: Multicycle PENVO with Delay}
\label{sec:floweqns}
On imposing parametric excitation to the nonlinearity in Eq.~(\ref{eq:kaiser_delay}), we can write,
\begin{eqnarray}
&&\ddot{x}+\mu \left[1+\gamma \cos(\Omega t)\right] \left( -1+x^2-\alpha x^4+\beta x^6   \right) \dot{x}\nonumber\qquad\quad\\
&&\phantom{\left[1+\gamma \cos(\Omega t)\right]1+x^2- x^4+\beta }+x- K x(t-\tau)=0.\qquad
\label{eq:kaiser_delay_pe}
\end{eqnarray}
The corresponding amplitude and phase equations are
\begin{subequations}
\begin{eqnarray}
\dot{\overline{r}}&=& -\frac{1}{128} \overline{r} \left(64 K \sin \tau+\mu  \left(5 \beta  \overline{r}^6-8 \alpha  \overline{r}^4+16 \overline{r}^2-64\right)\right)\nonumber\\
&&+A_\Omega(\overline{r},\overline{\phi};\gamma)+O(\mu^2);\\
\dot{\overline{\phi}}&=&-\frac{1}{2} K \cos \tau+B_\Omega(\overline{r},\overline{\phi};\gamma)+ O(\mu^2),
\end{eqnarray}
\label{eq:kaiser_delay_pe_amp_ph}
\end{subequations}
where higher order terms have been neglected, and $A_\Omega$ and $B_\Omega$ are functions with singularities at $\Omega= 2,\,4,\,6$ and $8$. One may resort to the L'H\^ospitals' rule and go to $p$-$q$ plane to rewrite the amplitude and the phase equations in terms of the coordinate of the plane:
\footnotesize
\begin{widetext}
\begin{eqnarray*}
\dot{p_2}&=&-\frac{K p \sin \tau}{2  }+\frac{K q \cos \tau}{2  }-\frac{1}{64} \beta  \gamma  \mu  p^7-\frac{1}{128} 5 \beta  \mu  p^7+\frac{1}{64} \alpha  \gamma  \mu  p^5+\frac{1}{16} \alpha  \mu  p^5+\frac{3}{32} \beta  \gamma  \mu  p^5 q^2-\frac{15}{128} \beta  \mu  p^5 q^2-\frac{\mu  p^3}{8}+\frac{15}{64} \beta  \gamma  \mu  p^3 q^4\\
&&-\frac{15}{128} \beta  \mu  p^3 q^4-\frac{5}{32} \alpha  \gamma  \mu  p^3 q^2+\frac{1}{8} \alpha  \mu  p^3 q^2-\frac{\gamma  \mu  p}{4}+\frac{\mu  p}{2}+\frac{1}{8} \beta  \gamma  \mu  p q^6-\frac{5}{128} \beta  \mu  p q^6-\frac{11}{64} \alpha  \gamma  \mu  p q^4+\frac{1}{16} \alpha  \mu  p q^4+\frac{1}{4} \gamma  \mu  p q^2-\frac{1}{8} \mu  p q^2,\\
\dot{q_2}&=&-\frac{K p \cos \tau}{2  }-\frac{K q \sin \tau}{2  }-\frac{1}{8} \beta  \gamma  \mu  p^6 q-\frac{5}{128} \beta  \mu  p^6 q-\frac{15}{64} \beta  \gamma  \mu  p^4 q^3-\frac{15}{128} \beta  \mu  p^4 q^3+\frac{11}{64} \alpha  \gamma  \mu  p^4 q+\frac{1}{16} \alpha  \mu  p^4 q-\frac{3}{32} \beta  \gamma  \mu  p^2 q^5\\
&&-\frac{15}{128} \beta  \mu  p^2 q^5+\frac{5}{32} \alpha  \gamma  \mu  p^2 q^3+\frac{1}{8} \alpha  \mu  p^2 q^3-\frac{1}{4} \gamma  \mu  p^2 q-\frac{1}{8} \mu  p^2 q+\frac{1}{64} \beta  \gamma  \mu  q^7-\frac{1}{128} 5 \beta  \mu  q^7-\frac{1}{64} \alpha  \gamma  \mu  q^5+\frac{1}{16} \alpha  \mu  q^5-\frac{\mu  q^3}{8}+\frac{\gamma  \mu  q}{4}+\frac{\mu  q}{2};\\
\dot{p_4}&=&-\frac{K p \sin \tau}{2  }+\frac{K q \cos \tau}{2  }+\frac{1}{64} \beta  \gamma  \mu  p^7-\frac{1}{128} 5 \beta  \mu  p^7-\frac{1}{32} \alpha  \gamma  \mu  p^5+\frac{1}{16} \alpha  \mu  p^5+\frac{9}{64} \beta  \gamma  \mu  p^5 q^2-\frac{15}{128} \beta  \mu  p^5 q^2+\frac{1}{16} \gamma  \mu  p^3-\frac{\mu  p^3}{8}\\&&
-\frac{5}{64} \beta  \gamma  \mu  p^3 q^4-\frac{15}{128} \beta  \mu  p^3 q^4-\frac{1}{16} \alpha  \gamma  \mu  p^3 q^2+\frac{1}{8} \alpha  \mu  p^3 q^2+\frac{\mu  p}{2}-\frac{13}{64} \beta  \gamma  \mu  p q^6-\frac{5}{128} \beta  \mu  p q^6+\frac{7}{32} \alpha  \gamma  \mu  p q^4+\frac{1}{16} \alpha  \mu  p q^4-\frac{3}{16} \gamma  \mu  p q^2-\frac{1}{8} \mu  p q^2,\\
\dot{q_4}&=&-\frac{K p \cos \tau}{2  }-\frac{K q \sin \tau}{2  }-\frac{13}{64} \beta  \gamma  \mu  p^6 q-\frac{5}{128} \beta  \mu  p^6 q-\frac{5}{64} \beta  \gamma  \mu  p^4 q^3-\frac{15}{128} \beta  \mu  p^4 q^3+\frac{7}{32} \alpha  \gamma  \mu  p^4 q+\frac{1}{16} \alpha  \mu  p^4 q+\frac{9}{64} \beta  \gamma  \mu  p^2 q^5\\
&&-\frac{15}{128} \beta  \mu  p^2 q^5-\frac{1}{16} \alpha  \gamma  \mu  p^2 q^3+\frac{1}{8} \alpha  \mu  p^2 q^3-\frac{3}{16} \gamma  \mu  p^2 q-\frac{1}{8} \mu  p^2 q+\frac{1}{64} \beta  \gamma  \mu  q^7-\frac{1}{128} 5 \beta  \mu  q^7-\frac{1}{32} \alpha  \gamma  \mu  q^5+\frac{1}{16} \alpha  \mu  q^5+\frac{1}{16} \gamma  \mu  q^3-\frac{\mu  q^3}{8}+\frac{\mu  q}{2};\\
\dot{p_6}&=&-\frac{K p \sin \tau}{2  }+\frac{K q \cos \tau}{2  }+\frac{1}{64} \beta  \gamma  \mu  p^7-\frac{1}{128} 5 \beta  \mu  p^7-\frac{1}{64} \alpha  \gamma  \mu  p^5+\frac{1}{16} \alpha  \mu  p^5-\frac{3}{32} \beta  \gamma  \mu  p^5 q^2-\frac{15}{128} \beta  \mu  p^5 q^2\\
&&-\frac{\mu  p^3}{8}-\frac{15}{64} \beta  \gamma  \mu  p^3 q^4-\frac{15}{128} \beta  \mu  p^3 q^4+\frac{5}{32} \alpha  \gamma  \mu  p^3 q^2+\frac{1}{8} \alpha  \mu  p^3 q^2+\frac{\mu  p}{2}+\frac{1}{8} \beta  \gamma  \mu  p q^6-\frac{5}{128} \beta  \mu  p q^6-\frac{5}{64} \alpha  \gamma  \mu  p q^4+\frac{1}{16} \alpha  \mu  p q^4-\frac{1}{8} \mu  p q^2,\\
\dot{q_6}&=&-\frac{K p \cos \tau}{2  }-\frac{K q \sin \tau}{2  }-\frac{1}{8} \beta  \gamma  \mu  p^6 q-\frac{5}{128} \beta  \mu  p^6 q+\frac{15}{64} \beta  \gamma  \mu  p^4 q^3-\frac{15}{128} \beta  \mu  p^4 q^3+\frac{5}{64} \alpha  \gamma  \mu  p^4 q+\frac{1}{16} \alpha  \mu  p^4 q\\
&&+\frac{3}{32} \beta  \gamma  \mu  p^2 q^5-\frac{15}{128} \beta  \mu  p^2 q^5-\frac{5}{32} \alpha  \gamma  \mu  p^2 q^3+\frac{1}{8} \alpha  \mu  p^2 q^3-\frac{1}{8} \mu  p^2 q-\frac{1}{64} \beta  \gamma  \mu  q^7-\frac{1}{128} 5 \beta  \mu  q^7+\frac{1}{64} \alpha  \gamma  \mu  q^5+\frac{1}{16} \alpha  \mu  q^5-\frac{\mu  q^3}{8}+\frac{\mu  q}{2};\\
\dot{p_8}&=&-\frac{K p \sin \tau}{2  }+\frac{K q \cos \tau}{2  }+\frac{1}{256} \beta  \gamma  \mu  p^7-\frac{1}{128} 5 \beta  \mu  p^7+\frac{1}{16} \alpha  \mu  p^5-\frac{21}{256} \beta  \gamma  \mu  p^5 q^2-\frac{15}{128} \beta  \mu  p^5 q^2-\frac{\mu  p^3}{8}\\
&&+\frac{35}{256} \beta  \gamma  \mu  p^3 q^4-\frac{15}{128} \beta  \mu  p^3 q^4+\frac{1}{8} \alpha  \mu  p^3 q^2+\frac{\mu  p}{2}-\frac{7}{256} \beta  \gamma  \mu  p q^6-\frac{5}{128} \beta  \mu  p q^6+\frac{1}{16} \alpha  \mu  p q^4-\frac{1}{8} \mu  p q^2,\\
\dot{q_8}&=&-\frac{K p \cos \tau}{2  }-\frac{K q \sin \tau}{2  }-\frac{7}{256} \beta  \gamma  \mu  p^6 q-\frac{5}{128} \beta  \mu  p^6 q+\frac{35}{256} \beta  \gamma  \mu  p^4 q^3-\frac{15}{128} \beta  \mu  p^4 q^3+\frac{1}{16} \alpha  \mu  p^4 q\\
&&-\frac{21}{256} \beta  \gamma  \mu  p^2 q^5-\frac{15}{128} \beta  \mu  p^2 q^5+\frac{1}{8} \alpha  \mu  p^2 q^3-\frac{1}{8} \mu  p^2 q+\frac{1}{256} \beta  \gamma  \mu  q^7-\frac{1}{128} 5 \beta  \mu  q^7+\frac{1}{16} \alpha  \mu  q^5-\frac{\mu  q^3}{8}+\frac{\mu  q}{2}.
\end{eqnarray*}
\end{widetext}
}
The subscript indicates the value of $\Omega$ in Eq.~(\ref{eq:kaiser_delay_pe}) for which the pair of above first order equations are written in ($p$,$q$) coordinates.

\bibliography{Saha_etal_bibliography}

\end{document}